\documentclass[12pt,a4paper]{article}

\usepackage{amsmath,upgreek}
\usepackage{amssymb}
\usepackage{mathrsfs}
\usepackage{framed}
\usepackage[colorlinks]{hyperref}
\usepackage{here}
\usepackage{graphicx}
\usepackage[dvipsnames]{xcolor}
\graphicspath{{./input/}}

\newcommand{\rem}[1]{}
\newcommand{\de}{{\rm d}}
\newcommand{\bq}{{\boldsymbol{q}}}
\newcommand{\bP}{{\mathbf{P}}}
\newcommand{\bp}{{\boldsymbol{p}}}
\newcommand{\bm}{{\mathbf{m}}}
\newcommand{\bX}{{\mathbf{X}}}

\newcommand{\bx}{{\boldsymbol{x}}}
\newcommand{\bn}{{\mathbf{n}}}

\newcommand{\bGamma}{{\boldsymbol{\Gamma}}}
\newcommand{\bsigma}{{\boldsymbol{\sigma}}}
\newcommand{\bz}{{\mathbf{z}}}
\newcommand{\bA}{\mathbf{A}}
\newcommand{\bE}{{\mathbf{E}}}

\newcommand{\bL}{{\boldsymbol{L}}}
\newcommand{\bu}{{\boldsymbol{u}}}
\newcommand{\bbeta}{{\boldsymbol{\beta}}}
\newcommand{\bS}{{\boldsymbol{S}}}
\newcommand{\bfi}{\bfseries\itshape}
\newcommand{\bmu}{\boldsymbol{\mu}}

\newcommand{\comment}[1]{\vspace{5mm}\par
\noindent
\framebox{\begin{minipage}[c]{.95\textwidth} \tt\bfi #1
\end{minipage}}\vspace{5 mm}\par}

\newcommand{\beq}{\begin{equation}}
\newcommand{\eeq}{\end{equation}}

\begin{document}

\title{
	Quantum-classical solvation hydrodynamics:\\a Hamiltonian modeling framework
}

\author{
	Fran\c{c}ois Gay-Balmaz$^1$ and Cesare Tronci$^{1,2}$
	\smallskip\\
	\small
	$^1$\it School of Physical and Mathematical Sciences, Nanyang Technological University, Singapore\\
	\small
	$^2$\it School of Mathematics and Physics, University of Surrey, Guildford, United Kingdom
}
\date{}

\maketitle
\begin{abstract}
We propose a mixed quantum-classical hydrodynamic framework to model short-time inertial effects in the non-adiabatic evolution of a quantum solute coupled to a classical polar solvent. Drawing upon the work of Burghardt and Bagchi \cite{BuBa06}, we employ the Hamiltonian approach to incorporate consistent backreaction and preserve quantum decoherence beyond standard Ehrenfest dynamics. The solvent is treated as an ideal polar fluid and the quantum solute state is coupled to both the  position and molecular orientation coordinates of the liquid. This approach retains essential solute-solvent correlations while  {reducing} the computational complexity of previous approaches. We further incorporate dissipative terms to capture both inertial effects and polarization relaxation.  After establishing the general setting for non-local dielectric continua, the Marcus local approximation is integrated into the model thereby extending traditional solvation theory to account for collective fluid sloshing on fast timescales.

\end{abstract}

\vspace{-.75cm}

{\footnotesize
\renewcommand{\contentsname}{} 
\tableofcontents
}

\section{Introduction}

\subsection{Modeling platforms in solvation dynamics\label{sec:intro}}

In solvation dynamics, the motion of the solute molecule is influenced by a dynamically evolving fluid solvent and this fluid-molecule interaction is critical to understand several processes
in both nature and technology. Given the interplay of several  nonlinear phenomena at different scales, many powerful methods have been devised over the years \cite{SaEtAl21}. These range from atomistic approaches based on classical molecular-mechanics (MM) force-fields or advanced QM/MM treatments \cite{SeTh09} to mesoscopic continuum models \cite{KlSc93,ToMeCa05} that typically provide only static descriptions. While the latter have recently been extended to the dynamical regime \cite{CaInMeTo05}, nonadiabatic decoherence effects remain elusive. Likewise, the accompanying solvent inertial response revealed by modern time-resolved ultrafast experiments needs to be incorporated, thereby going beyond the applicability of overdamped models \cite{DMPa20}.

In addressing the challenges above, a promising nonequilibrium
approach comprises hydrodynamic models \cite{BaJa10, BoEtAl11, BuBa06, ChBa91a,HuEtAl12, KiNi86,MaKa99}.  A remarkable effort in this direction is found in  \cite{BoEtAl11,BuBa06}, where the authors formulate a quantum-classical hydrodynamic model by applying the moment method  to the quantum-classical Liouville equation \cite{Aleksandrov,Gerasimenko,Kapral} for the hybrid density operator $\widehat{\cal P}(\bq,\bp)$ on phase-space. Depending on the closure adopted for the operator-valued stress tensor $\widehat{\Pi}$, the system presented in \cite{BoEtAl11} for the simple case of nonpolar solvation reads
\beq
\partial_t\hat{\boldsymbol{g}}   = -\operatorname{div}\widehat{\Pi}
-\frac1{2}\big(\tilde{\rho}(\nabla\widehat{H}_\textit{eff})+(\nabla\widehat{H}_\textit{eff})\tilde{\rho}\big)+\frac{1}{i\hbar} \big[\hat{\boldsymbol{g}},\widehat{H}_\textit{eff}\big]
,\qquad\qquad
     i\hbar \partial_t \tilde{\rho} +\frac{ i\hbar }M \operatorname{div}\hat{\boldsymbol{g}} =  \big[\widehat{H}_\textit{eff},\tilde{\rho}\big].
\label{burghmod}     
\eeq
Here,  $\tilde{\rho}$ and $\hat{\boldsymbol{g}}$  are  operator-valued density and momentum-density, respectively, so that the fluid  density and momentum  are    $D=\operatorname{Tr}\tilde{\rho}$ and $\bm=\operatorname{Tr}\hat{\boldsymbol{g}}$, {respectively, while the quantum state  reads $\hat\uprho=\int \tilde{\rho}\de^3q$}. In terms of the hybrid phase-space density $\widehat{\cal P}(\bq,\bp)$,  {the hydrodynamic variables} are given by the moments $\tilde{\rho}=\int \widehat{\cal P}\de^3p$ and $\hat{\boldsymbol{g}}=\int \bp \widehat{\cal P}\de^3p$. Also,  $\widehat{H}_\textit{eff}$ is an effective Hamiltonian carrying nonlocal terms due to the   molecular self-correlations of the solvent. 
Benchmarked in \cite{HuEtAl12} for certain closure schemes,
  the system \eqref{burghmod} provides a  promising perspective in solvation models beyond the current mainstream  in computational chemistry. An alternative hydrodynamic treatment based on Mori-Zwanzig projection operators also appeared recently in \cite{ScKa25}.
  In addition, polarization effects require a complex nonlinear treatment extending the usual fluid description to
incorporate the dipolar micromotion \cite{BaBi99}. While such complex fluid systems have a long
history, their coupling to quantum systems remains challenging. A remarkable step in this direction was made in  \cite{BuBa06,ChBa91a}, {using quantum-classical and fully classical formulations}.  

In this context, the quantum-classical coupling requires special attention since conventional approaches are affected by important limitations. Particular examples  are given by the violation of the Heisenberg principle  and the appearance of negative probabilities in configuration space. {While the latter have been studied in detail for surface hopping simulations \cite{Bondarenko} and quantum-classical Liouville implementations \cite{GuSc26},  popular mapping methods \cite{Richardson,Liu,Stark} also fail to guarantee positivity of the density matrix at all times.}  Other approaches such as Ehrenfest dynamics are well-known to struggle in capturing decoherence effects \cite{AkLoPr14}. Thus, in the search for effective hydrodynamic models of solvation dynamics, it is important to ensure that these limitations are overcome in the new setting, {while mitigating the computational complexity associated to nonlinear multiscale solvation processes.}

To provide a flexible modeling platform retaining frame-indifference,
energy balance, and positive probability, here we blend the Hamiltonian theory of  hydrodynamic fluid flows 
\cite{Ho11,HoMaRa98} with recent    quantum-classical models based on  Koopman wavefuctions in classical mechanics \cite{Koopman}. As initially proposed by Sudarshan \cite{Sudarshan}, these wavefunctions provide a Hilbert-space formulation of classical mechanics that is particularly suitable to blend the classical Liouville equation and the quantum Schr\"odinger equation. Along the way, we will adopt structure-preserving closures \cite{Morrison} and approximations within the total energy so that that the resulting model inherits a  Hamiltonian structure that is compatible with that of the original model. The fluid transport terms and the physical action/reaction forces   will be encoded  by the total energy, that is the Hamiltonian functional of the overall solvent-solute system. This approach is complementary to the conventional methodology, which is  based on physics-driven { closures and} approximations performed directly on the equations of motion, potentially compromising the fundamental consistency of the original system. {Indeed, this consistency is often  lost  by seemingly small steps, which no matter how simple, may inexorably destroy basic properties such as energy balance or the Heisenberg principle.} By a convenient definition of the dynamical variables, the methodology proposed here { also} allows to alleviate the computational complexity of previous hydrodynamic treatments.
On the one hand, this approach establishes a
rigorous Hamiltonian baseline in the ideal inviscid solvent limit, providing a consistent foundation before the introduction of dissipative effects.
On the other hand, it goes beyond  quantum-classical Ehrenfest dynamics by capturing decoherence while retaining the Heisenberg principle \cite{BaBeGBTr24}. Dissipation and irreversible effects are added \emph{a posteriori} by adopting  tools from dynamical density-functional theory \cite{ChBa91b,GoNoKa16,Vrugt}. For example, following the treatment in \cite{BoEtAl11}, one can model dissipation by  adding a suitable Stokes drag term as well as shear and bulk viscosity. 
Then, the overall purpose of this paper is to provide a general modeling setting that can unfold the conservative and dissipative mechanisms underlying solute decoherence and solvent inertial response, beyond the limitations affecting current { treatments}. 
From a chemical perspective, the structural consistency afforded by this Hamiltonian-based approach provides a rigorous foundation for characterizing the ultrafast interfacial dynamics that govern molecular processes. By establishing a first-principles dynamical baseline for an ideal fluid solvent, this framework allows for the systematic isolation of inertial and polarization effects—often obscured by empirical damping—on the solute's electronic state.  Consequently, this model offers a fundamental platform to investigate the role of solvent-induced decoherence in critical chemical phenomena such as charge-transfer mechanisms in complex environments.
By ensuring that the underlying conservation laws are strictly satisfied in the ideal limit, along with stringent consistency criteria, the present theory facilitates the development of robust, energy-consistent benchmarks that are essential for the next generation of predictive multiscale simulations in computational chemistry.  {In addition, comparing to previous models, a lower dimensionality in both the number of time-evolving fields and the chosen  coordinate space leads to a relief in computational complexity.}

\paragraph{Plan of the paper.}
The next section proceeds by presenting explicitly the  Koopman model for quantum-classical dynamics on phase-space while illustrating its energy functional and Hamiltonian structure. {Subsequently}, suitable hydrodynamic closures are presented in \S\ref{sec:MQCHydro1} for the simple case of nonpolar solvation. After applying in \S\ref{sec:MQCHydroEhr} the Hamiltonian closure method to Ehrenfest hydrodynamics for barotropic equations of state, the treatment is extended in \S\ref{sec:MQCHydroAdbEhr} to account for entropy transport in adiabatic flows. Later on, the discussion continues by considering the closure of backreaction terms in the hydrodynamic Koopman model. The Poisson bracket structure is presented throughout and is used as a guidance for the Hamiltonian closure approach developed in the paper. In \S\ref{sec:NOEx}, the Koopman hydrodynamic model is then specialized to the case of two-level electronic transitions of a nitric oxide molecule in supercritical argon. The treatment is extended to account for the solvent polarization in \S\ref{sec:PolSolv}. Unlike previous approaches, here the hydrodynamic variables are still defined on physical space, while the solute  state remains correlated to both the position and orientation coordinates of the solvent.  As a result, the approach presented here allows to alleviate computational complexity.  Once again, the treatment first focuses on the Ehrenfest model, which is presented in \S\ref{sec:MicEhr}. Later, a hybrid Ehrenfest-Koopman picture is presented in \S\ref{sec:hybPolKoop}, where the orientational backreaction is included within the Koopman model, while the backreaction associated to translational coordinates remains treated at the Ehrenfest level. In \S\ref{MarcusHyd}, this hybrid model is specialized to consider nonlocal interactions in dielectric solvents. Within the same section, the computational complexity is further alleviated by applying Marcus' local approximation, thereby obtaining an ideal model referred to as \emph{Marcus hydrodynamics}. Finally, in \S\ref{sec:dissip} the presentation concludes not only by restoring conventional dissipative effects such as friction and viscosity but also by devising a new diffusive term to render polarization diffusion within the  equation of motion of the entire  quantum solute state. Inspired by analogies with dynamical DFT, this term only affects the solvent orientational dynamics while retaining the unitary nature of the solute evolution.

\subsection{The quantum-classical Koopman model on phase-space}

Following the  approach adopted by Burghardt and coauthors \cite{BoEtAl11}, we will devise our hydrodynamic models by applying the moment method from kinetic theory within the mixed quantum-classical setting. However, instead of focusing on the quantum-classical Liouville equation \cite{Aleksandrov,Gerasimenko,Kapral}, we will take the kinetic moments of a hybrid density operator $\widehat{\cal P}(\bq,\bp)$ obeying a nonlinear quantum-classical equation  recently introduced by the authors as the \emph{Koopman model} \cite{BaBeGBTr24,GBTr22,GBTr21}. 
In this context, the hybrid density is expressed as a projection operator $\widehat{\cal P}(\bq,\bp,\bx,\bx')=\Upsilon(\bq,\bp,\bx)\Upsilon(\bq,\bp,\bx')^*$, where $\Upsilon(\bq,\bp,\bx)$ is a quantum-classical wavefunction in the tensor product space of Koopman wavefunctions on the classical phase-space and Schr\"odinger wavefunctions in the quantum configuration space. 
A Koopman wavefunction $\chi(\bq,\bp,t)$ in classical mechanics evolves according to $i\hbar\partial_t\chi=i\hbar\{H,\chi\}+\phi$, where $\phi(\bq,\bp)$ is an arbitrary phase function, so that $|\chi|^2$ evolves according to the Liouville equation $\partial_t|\chi|^2=\{H,|\chi|^2\}$. Here, the operator $i\hbar\{H,\cdot\,\}$ is self-adjoint, that is $\chi$ evolves under linear unitary dynamics thereby establishing a direct analogy with quantum mechanics. {In this context, statistical properties transfer directly from the standard Liouville picture  by simply using the relation ${\cal D}=|\chi|^2$, where ${\cal D}$ is the classical Liouville density. Then,  Koopman wavefunctions provide an alternative and yet equivalent picture of classical Liouville dynamics}.

Initially formulated as a linear \emph{quantum-classical wave equation} for the hybrid wavefunction $\Upsilon(\bq,\bp,\bx)$ \cite{BoGBTr19,GBTr20,Manfredi23}, the Koopman model was taken into its current nonlinear form   by performing further modeling steps to guarantee positivity in both quantum and classical sectors. These steps were presented and further summarized in a series of recent papers \cite{BaBeGBTr24,GBTr22,GBTr21} and essentially consist in applying a symmetry principle that, inspired by Sudarshan's work \cite{Ghose,Sudarshan}, makes classical phase functions into an arbitrary unobservable gauge. {Without this gauge principle, the  quantum-classical wave equation previously proposed in \cite{BoGBTr19} retains positivity of the quantum density matrix, while a similar result for the classical density has so far only been proved for pure-dephasing dynamics \cite{GBTr20}. In addition, Sudarshan's suggestions to make classical phases unobservable is very grounded in physics: we do not measure Hamilton-Jacobi functions. Therefore, here we will focus on the nonlinear model variant proposed in \cite{BaBeGBTr24,GBTr22,GBTr21}, which intrinsically satisfies several physical consistency requirements, including positivity of both quantum and classical probabilities.

In  its Hamiltonian form}, the quantum-classical Koopman model reads \cite{BaBeGBTr24,GBTr21}
\beq\label{KMod1}
\frac{\partial \widehat{\cal P}}{\partial t}
+{\operatorname{div}}\big(\widehat{\cal P}\big\langle\bX_{\delta h/\delta\widehat{\cal P}}\big\rangle\big)
=-\frac{i}\hbar\left[\frac{\delta h}{\delta\widehat{\cal P}},\widehat{\cal P}\right]\!
,\quad\ \ \, \text{where}\quad\ \ \,
h(\widehat{\cal P})={\int}\big\langle\widehat{H}{\operatorname{Tr}}\widehat{\cal P}+i\hbar\{\widehat{\cal P},\widehat{H}\}\big\rangle \de^3q \de^3p
\eeq
is the total energy, i.e. the Hamiltonian functional of the system. The notation is such that $\widehat{H}(\bq,\bp)$ is the quantum-classical Hamiltonian,  $\bX_{\widehat{A}}=(\partial_\bp\widehat{A},-\partial_\bq\widehat{A})$ is the phase-space Hamiltonian vector field associated to the quantum-classical observable $\widehat{A}(\bq,\bp)$. Also, $\operatorname{Tr}$ is the trace associated to the quantum state space and 
\[
\langle\widehat{A}\rangle=\frac{{\operatorname{Re}}{\operatorname{Tr}}(\widehat{\cal P}\widehat{A})}{\operatorname{Tr}\widehat{\cal P}}
\] 
denotes the local expectation. Before presenting the explicit form of  equation \eqref{KMod1}, we observe that the energy density comprises two terms: the first is a standard expectation corresponding to the usual Ehrenfest dynamics; the second carries phase-space inhomogeneities that are responsible for the \emph{quantum backreaction} and lead to decoherence effects. In this sense, the Koopman model can be understood as an $\hbar$-correction of the usual Ehrenfest model   \cite{Vanicek}
\beq\label{EhrPSeqn}
{i}\hbar\frac{\partial \widehat{\cal P}}{\partial t}
+{i}\hbar{\operatorname{div}}(\widehat{\cal P}\langle\bX_{\widehat{H}}\rangle)
=[\widehat{H},\widehat{\cal P}].
\eeq

As shown in Appendix \ref{sec:Appx}, upon denoting the classical density by ${\cal D}={\operatorname{Tr}}\widehat{\cal P}$, the explicit form of equation \eqref{KMod1} reads
\beq\label{KMod2}
i\hbar\frac{\partial \widehat{\cal P}}{\partial t}+i\hbar{\,\operatorname{div}}\bigg(\frac{\widehat{\cal P}}{{\cal D}}{\operatorname{Tr}}\big[\bX_{\widehat{H}},\widehat\bGamma\big]_\text{\!\bfi JL}+(\widehat{\cal P}\bX_{\widehat{H}})^{\sf H}\bigg)=\big[\widehat{H},\widehat{\cal P}\big]+\big[\widehat{\bGamma},\nabla{\widehat{H}}\big].
\eeq
Here, the notation is as follows:
\beq\label{somedefs}
\widehat{A}^{\,\sf H}=\frac12(\widehat{A}+\widehat{A}^\dagger)
,\qquad\qquad
[{\bf A},{\bf B}]_\text{\bfi JL}={\bf A}\cdot\nabla{\bf B}-{\bf B}\cdot\nabla{\bf A}
,\qquad\qquad
\widehat\bGamma
=\frac{i\hbar}{2{\cal D}}\big[\widehat{\cal P},\bX_{\widehat{\cal P}}\big], 
\eeq
so that $[\cdot,\cdot]_\text{\bfi JL}$
denotes the \emph{Jacobi-Lie commutator} on vector fields, while the quantity $\widehat\bGamma$  incorporates inhomogeneities of the quantum-classical state. A similar quantity also appears in molecular geometric phases \cite{Me92} and spin-orbit coupling \cite{Tronci26}, thereby leading to interesting analogies. The total energy is rewritten in terms of $\widehat\bGamma$  as $h(\widehat{\cal P})={\operatorname{Tr}}{\int}\widehat{\mathscr{D}}\widehat{H} \de^3q \de^3p$, where the pseudo-density operator  $\widehat{\mathscr{D}}=\widehat{\cal P}+\operatorname{div}\widehat\bGamma$ enjoys important covariance properties summarized in \cite{GBTr21}. For more details on pseudo-density operators, see \cite{Fullwood}.

Benchmarked against the nonadiabatic Tully problems \cite{BaBeGBTr24,Tully90} and the Rashba dynamics of quantum nanowires \cite{BeMaTr26} via a numerical particle scheme, the Koopman model \eqref{KMod2} was recently shown not only to retain positivity at all times but also to capture quantum decoherence with accuracy levels well beyond Ehrenfest dynamics. In addition, despite its formidable look, equation \eqref{KMod2} shows an important relation with the quantum-classical Liouville equation \cite{Kapral}, which is indeed recovered upon discarding all $\widehat\bGamma$-contributions. However, we point out that in this limit the positive-definite projection solutions $\widehat{\cal P}=\Upsilon\Upsilon^\dagger$ are no longer available and one is forced to enlarge the solution space to consider general Hermitian operators. As already noticed in \cite{TrGB23}, the Koopman model in \eqref{KMod2} does not involve gradients of order higher than 2 and its geometric variational and Hamiltonian structures have recently led to various closure schemes for further modeling and numerical implementation. Indeed, 
the model in \eqref{KMod2} has been studied from different perspectives. The associated Heisenberg picture was recently presented in \cite{DMCTr24}, while the extension to quantum-classical spin systems is found in \cite{GBTr23}. The entropy functional associated to both Ehrenfest and Koopman models was introduced in \cite{TrMCGB}, along with the maximal-entropy equilibrium states. A particle scheme implementation and a hydrodynamic fluid closure were presented in \cite{BaBeGBTr24} and \cite{GBTr-fluid}, respectively.

Given the level of complexity of the continuum equation \eqref{KMod2}, one is led to go beyond standard methodologies in studying the properties of the Koopman model. In particular, variational  action principles have proved very successful and all the studies so far have been based on the variational approach. The underlying idea is to perform closures and approximations at the level of the variational principle underlying \eqref{KMod2} in such a way to retain its  mathematical structure. However, the variational principle associated to \eqref{KMod2} is not based on a straightforward Dirac-Frenkel-type action and rather resorts to the Euler-Poincar\'e theory of reduction from Lagrangian to Eulerian variables in continuum mechanics \cite{Ho11,HoMaRa98}. While Euler-Poincar\'e variational principles have proved very effective in a variety of different problems, they are based on constrained variations with Lin-type constraints \cite{CeMa87} that require special care. In order to avoid this type of complication, here we propose an alternative approach in which the entire modeling approach is based on the Hamiltonian form \eqref{KMod1} of equation \eqref{KMod2}. In the theory of mechanical systems, a Hamiltonian structure is provided in terms of two objects: a Hamiltonian (total energy) functional and a Poisson bracket. Without going through the details, here we will simply say that the Poisson bracket satisfies the axioms of a Lie bracket as well as the Leibniz product rule from standard differential calculus.  As discussed in \cite{GBTr21,TrGB23}, the mixed quantum-classical (MQC) Poisson bracket associated to \eqref{KMod1} is given by
\beq\label{Koopbkt}
 \{f,k\}_\text{MQC}={\operatorname{Tr}}{\int}\widehat{\mathcal{P}}\left(\left\langle\nabla \frac{\delta f}{\delta \widehat{\mathcal{P}}}\right\rangle\cdot\Bbb{J}\left\langle\nabla \frac{\delta k}{\delta \widehat{\mathcal{P}}}\right\rangle -\frac{i}\hbar\left[\frac{\delta f}{\delta \widehat{\mathcal{P}}},\frac{\delta k}{\delta \widehat{\mathcal{P}}}\right]\right)\de^3q\de^3p,
\eeq
where $\Bbb{J}^{jk}=\{z^j,z^k\}$ and we use the notation $\bz=(\bq,\bp)$ for points on phase-space. In agreement with the general properties of Poisson brackets, we observe that equation \eqref{KMod1} follows from the relation $\dot{f}= \{h,f\}_\text{MQC}$ holding for the Hamiltonian functional $h(\widehat{\mathcal{P}})$ in \eqref{KMod1} and any arbitrary functional $f=f(\widehat{\mathcal{P}})$. Hence,  operating on the Poisson bracket is equivalent to operating on the  Hamiltonian functional $h$ entering the generalized Hamiltonian equation \eqref{KMod1}. This is the approach  followed in this paper: we will operate closures and approximations at the level of the total energy and take further steps that are compatible with the  Hamiltonian structure  underlying the original Koopman model.

\section{Hybrid quantum-classical hydrodynamics\label{sec:MQCHydro1}}

To overcome the dimensional complexity of phase-space simulations, a hydrodynamic closure of the Koopman model  \eqref{KMod2} was recently formulated in \cite{GBTr-fluid}. Therein, a variational approach was devised in such a way that the resulting hydrodynamic equations inherit a Hamiltonian structure from the original phase-space model. Here, we will illustrate the Hamiltonian setting for this general methodology by proceeding  in two stages: first, we will simply derive the Ehrenfest hydrodynamic model by dropping the backreaction energy ${\int}\langle i\hbar\{\widehat{\cal P},\widehat{H}\}\rangle \de^3q \de^3p$ in \eqref{KMod1}; second, the latter will be restored and further closures will be adopted therein.

\subsection{Hamiltonian moment method for Ehrenfest hydrodynamics\label{sec:MQCHydroEhr}}

In order to apply the moment method in a Hamiltonian setting, we introduce the hydrodynamic moments
\beq\label{moms}
D={\operatorname{Tr}}{\int} \widehat{\cal P}\de^3p
\,,\qquad
\bm={\operatorname{Tr}}{\int}\bp \widehat{\cal P}\de^3p
\,,\qquad
\tilde\rho={\int} \widehat{\cal P}\de^3p.
\eeq
While the first two are the standard density and momentum density in configuration space, the latter is a hybrid distribution-valued density matrix retaining the information on the quantum state. We notice that $D={\operatorname{Tr}}\tilde\rho$.
Let us now assume that the Hamiltonian functional $h(\widehat{\mathcal{P}})$ in \eqref{KMod1} can be written in terms of the hydrodynamic quantities above. That is, we assume that there is a functional ${\sf h}(D,\bm,\tilde\rho)$ such that ${\sf h}(D,\bm,\tilde\rho)=h(\widehat{\mathcal{P}})$.  Instead of prescribing an explicit expression, at this stage we leave the new functional   ${\sf h}$  arbitrary so that the chain rule gives
\beq\label{funCR}
\frac{\delta h}{\delta \widehat{\mathcal{P}}}=\frac{\delta \sf h}{\delta D}+\bp\cdot\frac{\delta \sf h}{\delta \bm}+\frac{\delta \sf h}{\delta \tilde\rho},
\eeq
where we notice that all three functional derivatives on the right-hand side depend only on configuration space coordinates.
Replacing this relation in the generalized Hamiltonian equation from \eqref{KMod1} and taking moments yields the following hydrodynamic Hamiltonian equations:
\begin{align}\label{HHam1}
&\frac{\partial D}{\partial t}+\operatorname{div}\left(\frac{\delta \sf h}{\delta \bm}D\right)=0
\\\label{HHam2}
&\frac{\partial \tilde\rho}{\partial t}+\operatorname{div}\left(\frac{\delta \sf h}{\delta \bm}\tilde\rho\right)=-\frac{i}\hbar\left[\frac{\delta \sf h}{\delta \tilde\rho},\tilde\rho\right]
\\\label{HHam3}
&\frac{\partial \bm}{\partial t}+\operatorname{div}\left(\frac{\delta \sf h}{\delta \bm}\bm\right)+\nabla\frac{\delta \sf h}{\delta \bm}\cdot\bm=-D\nabla\frac{\delta \sf h}{\delta D}-{\operatorname{Tr}}\left(\tilde\rho\nabla\frac{\delta \sf h}{\delta \tilde\rho}\right)
\end{align}
At this point, the full set of hydrodynamic equations can be provided upon specifying the particular form of the total energy ${\sf h}(D,\bm,\tilde\rho)$. For example, let us consider the total energy $h(\widehat{\mathcal{P}})$ in \eqref{KMod1} with the quantum-classical Hamiltonian operator $\widehat{H}=M^{-1}p^2/2+\widehat{\cal H}(\bq)$ and perform the closure {replacement}
\beq\label{ClosHamEhr1}
{\operatorname{Tr}}{\int}\left(\frac{p^2}{2M}+\widehat{\cal H}\right)\widehat{\cal P} \de^3q \de^3p\simeq{\int}\left(\frac1{2M}\frac{m^2}{D}+D\mathscr{U}(D) + {\operatorname{Tr}}(\tilde\rho\widehat{\cal H})\right) \de^3q,
\eeq
where the first term in the second integral is the hydrodynamic kinetic energy and the second term involves a thermodynamic internal energy $\mathscr{U}(D)$ in the barotropic approximation, that is the internal energy depends only on the density $D$.  In the context of the Ehrenfest dynamics \eqref{EhrPSeqn}, we neglect the backreaction energy:
\[
{\int}\big\langle i\hbar\{\widehat{\cal P},\widehat{\cal H}\}\big\rangle \de^3q \de^3p\simeq0.
\]
Then, upon introducing 
\beq\label{urhovars}
\bu=\frac{\bm}{MD} \qquad \text{ and } \qquad \hat\rho=\frac{\tilde\rho}D,
\eeq
 the hydrodynamic Hamiltonian equations \eqref{HHam1}-\eqref{HHam3} yield the Ehrenfest hydrodynamic model
\begin{align}\label{EhrHam1}
&\frac{\partial D}{\partial t}+\operatorname{div}\left(\bu D\right)=0,
\\\label{EhrHam2}
&\frac{\partial \hat\rho}{\partial t}+\bu\cdot\nabla\hat\rho=-\frac{i}\hbar[\widehat{\cal H},\hat\rho],
\\\label{EhrHam3}
&\frac{\partial \bu}{\partial t}+\bu\cdot\nabla\bu=-\frac1{MD}\nabla{\sf p}-\frac1M{\operatorname{Tr}}(\hat\rho\nabla\widehat{\cal H}),
\end{align}
where the hydrodynamic pressure is given by ${\sf p}=D^2\mathscr{U}'(D)$. Equations \eqref{EhrHam1}-\eqref{EhrHam3} represent the most basic quantum-classical hydrodynamic model. While the Ehrenfest approach is known to fail in capturing decoherence, the model above serves as a mathematical modeling platform for adding different layers of complexity depending on the desired accuracy. Its Poisson structure is revealed explicitly upon evaluating the time derivative $\de{f}/\de t={\int}(\delta f/\delta D)\partial_t D\de^3 q+{\int}(\delta f/\delta\bm)\cdot\partial_t \bm\de^3 q+{\operatorname{Tr}}{\int}(\delta f/\delta\tilde\rho)\partial_t \tilde\rho\de^3q$ of an arbitrary functional  along the dynamics prescribed by \eqref{HHam1}-\eqref{HHam3}. Using the relation $\dot{f}=\{f,h\}_\text{H-MQC}$, we obtain the  hydrodynamic MQC bracket
\begin{align}\nonumber
\{f,k\}_\text{H-MQC}=&{\int}\bm\cdot\left(\frac{\delta k}{\delta \bm}\cdot\nabla\frac{\delta f}{\delta \bm}-\frac{\delta f}{\delta \bm}\cdot\nabla\frac{\delta k}{\delta \bm}\right)\de ^3 q
-
{\int}D\left(\frac{\delta f}{\delta \bm}\cdot\nabla\frac{\delta k}{\delta D}-\frac{\delta k}{\delta \bm}\cdot\nabla\frac{\delta f}{\delta D}\right)\de ^3 q
\nonumber
\\
&
-{\operatorname{Tr}}{\int} \tilde\rho\left( i\hbar^{-1}\left[\frac{\delta k}{\delta \tilde\rho},\frac{\delta f}{\delta \tilde\rho}\right]+\frac{\delta f}{\delta \bm}\cdot\nabla\frac{\delta k}{\delta \tilde\rho}-\frac{\delta k}{\delta \bm}\cdot\nabla\frac{\delta f}{\delta  \tilde\rho}\right)\de^3  q,
\label{Ehr-brkt}
\end{align}
which is equivalently found also by using the functional chain rule relation \eqref{funCR} in \eqref{Koopbkt}.

The Poisson bracket above also allows for the existence of the so-called \emph{Casimir invariants}. These are dynamical invariants  $C$ defined by the property $\{C,f\}_\text{H-MQC}=0$, for any functional $f$, so that $\dot{C}=\{C,h\}_\text{H-MQC}=0$ regardless of the form of $h$. An obvious Casimir invariant is given by 
\beq\label{Casimir1}
C_1=\int\! D \Phi(D^{-1}\tilde\rho)\de^3q, 
\eeq
where $\Phi$ is any real-valued function on the space of square complex matrices. For example, $\Phi(\widehat{A})=-{\operatorname{Tr}}(\widehat{A}\ln \widehat{A})$ recovers the usual prescription for the von Neumann entropy. Another Casimir is made available upon writing $\tilde\rho=D\psi\psi^\dagger$. In this case, the \emph{hydrodynamic helicity} invariant is given by \cite{GBTr-fluid}
\beq\label{Helicity}
C_2={\int}(M\bu-\bA)\cdot\nabla\times(M\bu-\bA)\,\de^3q
\,,\qquad\text{where}\qquad
\bA=-i\hbar\psi^\dagger\nabla\psi
\eeq
is the Berry connection \cite{Berry1984}. In the case that $\hat\rho=\tilde\rho/D$ is not a projection, one can resort to the Uhlmann operators \cite{Bondar,Tronci19,Uhlmann} to write $\tilde \rho=DWW^\dagger$, where $W$ is any rectangular matrix satisfying $\|W\|^2=1$, so that the Berry connection  is given by $\bA=-i\hbar\operatorname{Tr}(W^\dagger\nabla W)$. Casimir invariants are relevant in the study of nonlinear stability by Arnold's energy-Casimir method and its extension to continuum theories \cite{HoMaRaWe85}.

\subsection{Beyond barotropic hydrodynamics: the adiabatic closure\label{sec:MQCHydroAdbEhr}}

In the previous section we made the approximation that the thermodynamic internal energy depends only on the fluid density, that is $\mathscr{U}=\mathscr{U}(D)$. Examples are provided by politropic expressions with a fixed  exponent. However, more realistic situations involve the presence of a specific entropy $s$ so that the equation of state reads 
\[
\de\mathscr{U}=-{\sf p}\de(1/D)+T\de s,
\]
where ${\sf p}(D,s)$ and $T(D,s)$ are the pressure and local temperature, respectively. 

We notice that the specific entropy $s(\bq,t)$ emerges in this context as an extra variable whose dynamics must be suitably prescribed as part of the closure procedure. Since we are working in the framework of continuum Hamiltonian systems, we need to prescribe a dynamics of $s$ that is  compatible with the Hamiltonian structure of the barotropic case, in the sense that the latter must be recovered when the specific entropy is ignored. In the context of reversible hydrodynamics the specific entropy is evidently preserved along the flow, that is $\partial_t s+\bu\cdot\nabla s=0$. Within the Hamiltonian setting, {given the kinetic energy $M^{-1}\int (D^{-1}m^2/2)\de^3q$}, this means that the  system  \eqref{HHam1}-\eqref{HHam3} must be extended to include the equation
\beq\label{EntEq}
\frac{\partial s}{\partial t}+\frac{\delta \sf h}{\delta \bm}\cdot\nabla s=0.
\eeq
This procedure must  be completed by expressing the feedback force term that is produced by the entropy transport in the fluid momentum equation. This further step can be performed by resorting to the Poisson bracket structure. To understand how \eqref{Ehr-brkt} must be modified to account for entropy transport, let us expand the time derivative of an arbitrary functional $f=f(\bm,\tilde\rho,D,s)$ as 
\[
\frac{\de{f}}{\de t}={\int}\frac{\partial D}{\partial t}\frac{\delta f}{\delta D}\de^3 q+{\int}\frac{\partial \bm}{\partial t}\cdot\frac{\delta f}{\delta\bm}\de^3 q+{\operatorname{Tr}}{\int}\frac{\partial \tilde\rho}{\partial t}\frac{\delta f}{\delta\tilde\rho}\de^3q+{\int}\frac{\partial s}{\partial t}\frac{\delta f}{\delta s}\de^3 q
.
\]
 Using \eqref{EntEq}, the last term reads
\[
{\int}\frac{\partial s}{\partial t}\frac{\delta f}{\delta s}\de^3 q={\int}s\operatorname{div}\!\left(\frac{\delta \sf h}{\delta \bm}\frac{\delta f}{\delta s}\right)\de^3 q.
\]
Then, since $\dot{f}=\{f,h\}_\text{H-MQC}$, the above term must be added to the Poisson bracket \eqref{Ehr-brkt} to construct the extension for adiabatic fluid flows. {Also,} since the Poisson bracket must be antisymmetric, the above term must be accompanied by an additional analogous term with opposite sign, in which ${\sf h}$ and $f$ are swapped. Thus, we write the extension of \eqref{Ehr-brkt} to adiabatic flows as
\begin{align}\nonumber
\{f,k\}_\text{H-MQC}=&{\int}\bm\cdot\left(\frac{\delta k}{\delta \bm}\cdot\nabla\frac{\delta f}{\delta \bm}-\frac{\delta f}{\delta \bm}\cdot\nabla\frac{\delta k}{\delta \bm}\right)\de ^3 q
\\\nonumber
&
-{\operatorname{Tr}}{\int} \tilde\rho\left( i\hbar^{-1}\left[\frac{\delta k}{\delta \tilde\rho},\frac{\delta f}{\delta \tilde\rho}\right]+\frac{\delta f}{\delta \bm}\cdot\nabla\frac{\delta k}{\delta \tilde\rho}-\frac{\delta k}{\delta \bm}\cdot\nabla\frac{\delta f}{\delta  \tilde\rho}\right)\de^3  q
\\
&
-
{\int}D\left(\frac{\delta f}{\delta \bm}\cdot\nabla\frac{\delta k}{\delta D}-\frac{\delta k}{\delta \bm}\cdot\nabla\frac{\delta f}{\delta D}\right)\de ^3 q
-
\int \! s\operatorname{div}\!\left(\frac{\delta f}{\delta \bm}\frac{\delta k}{\delta s}-\frac{\delta k}{\delta \bm}\frac{\delta f}{\delta s}\right)\de ^3 q.
\label{Ehr-brkt-adb}
\end{align}
At this point, we should proceed with extra care: while we have ensured antisymmetry in constructing the last term above, the other defining properties of Poisson brackets deserve special attention. Notably, the Jacobi identity $\{\ell,\{f,k\}\}+\{k,\{\ell,f\}\}+\{f,\{k,\ell\}\}=0$ holding for any three arbitrary functionals $\ell$, $f$, and $k$ is the most difficult property to capture in Hamiltonian dynamics. For example, the bracket structure underlying the quantum-classical Liouville equation fails to retain this important property, thereby leading to well-known issues \cite{AgCi07,Sergi}. Luckily, in the present case we are dealing with the fluid transport of a scalar function as in \eqref{EntEq} so that the treatment falls within the class of \emph{Lie-Poisson systems}  \cite{MaRa98}, which are well known in the theory of Hamiltonian mechanics. In particular, both the brackets \eqref{Ehr-brkt} and \eqref{Ehr-brkt-adb} are Poisson brackets of Lie-Poisson type \cite{MaRaWe84}, thereby satisfying all the defining properties including the Jacobi identity. Notice that the new bracket \eqref{Ehr-brkt-adb} also allows for an extra family of Casimir invariants other than those already discussed in the previous section. Indeed, in the case $\tilde\rho=D\psi\psi^\dagger$, we have \eqref{Casimir1} as well as the family of invariants given by $C_3=\int D\Gamma(s,\Omega)\de^3 q$, where $\Gamma$ is an arbitrary function of two variables and $\Omega = D^{-1}\nabla s\cdot\nabla\times(M\bu-\bA)$ is the so-called \emph{potential vorticity}. On the other hand,  entropy transport prevents conservation of the hydrodynamic helicity \eqref{Helicity}.

At this point, the relation $\dot{f}=\{f,{\sf h}\}_\text{H-MQC}$ yields the generalized Hamiltonian equations for adiabatic Ehrenfest hydrodynamics:
\begin{align}\label{HHam1adi}
&\frac{\partial D}{\partial t}+\operatorname{div}\left(\frac{\delta \sf h}{\delta \bm}D\right)=0
\\\label{HHam2adi}
&\frac{\partial s}{\partial t}+\frac{\delta \sf h}{\delta \bm}\cdot\nabla s=0
\\\label{HHam3adi}
&\frac{\partial \tilde\rho}{\partial t}+\operatorname{div}\left(\frac{\delta \sf h}{\delta \bm}\tilde\rho\right)=-\frac{i}\hbar\left[\frac{\delta \sf h}{\delta \tilde\rho},\tilde\rho\right]
\\\label{HHam4adi}
&\frac{\partial \bm}{\partial t}+\operatorname{div}\left(\frac{\delta \sf h}{\delta \bm}\bm\right)+\nabla\frac{\delta \sf h}{\delta \bm}\cdot\bm=\frac{\delta \sf h}{\delta s}\nabla s-D\nabla\frac{\delta \sf h}{\delta D}-{\operatorname{Tr}}\left(\tilde\rho\nabla\frac{\delta \sf h}{\delta \tilde\rho}\right).
\end{align}
Then, taking the relevant functional derivatives of the total energy
\[
{\sf h}={\int}\left(\frac1{2M}\frac{m^2}{D}+D\mathscr{U}(D,s) + {\operatorname{Tr}}(\tilde\rho\widehat{\cal H})\right) \de^3q, 
\]
we obtain the adiabatic Ehrenfest equations
\begin{align}\label{EhrHam1-ad}
&\frac{\partial D}{\partial t}+\operatorname{div}\left(\bu D\right)=0,
\\\label{EhrHam2-ad}
&\frac{\partial s}{\partial t}+\bu \cdot\nabla s=0,
\\\label{EhrHam3-ad}
&\frac{\partial \hat\rho}{\partial t}+\bu\cdot\nabla\hat\rho=-\frac{i}\hbar[\widehat{\cal H},\hat\rho],
\\\label{EhrHam4-ad}
&\frac{\partial \bu}{\partial t}+\bu\cdot\nabla\bu=-\frac1{MD}\nabla{\sf p}-\frac1M{\operatorname{Tr}}(\hat\rho\nabla\widehat{\cal H}),
\end{align}
where the pressure is given as ${\sf p}(D,s)=D^2\partial\mathscr{U}/\partial D$.

The technique illustrated here to add the transport of scalar functions -- in this case, the specific entropy $s$ -- may be used as a systematic method by operating on the Poisson bracket structure. Whenever a Hamiltonian hydrodynamic system needs to be modified by the addition of a transported {scalar} function, one can proceed by adding to the original Poisson bracket an analogous term to the last in \eqref{Ehr-brkt-adb}. This procedure will be used throughout the reminder of this paper.

\subsection{Hydrodynamic moment closure of the  Koopman model\label{sec:KoopHyd1}}

We now move on to discuss the hydrodynamic closure of the mixed quantum-classical Koopman model \eqref{KMod2}. While such a closure was previously formulated in terms of the variational principle underlying the model \cite{GBTr-fluid}, here we want to proceed entirely within the Hamiltonian setting.

The first task is to devise a closure for the energy $h(\widehat{\cal P})$ in \eqref{KMod1} in such a way that it can be expressed in terms of the hydrodynamic variables \eqref{moms}, along with auxiliary variables that may be necessary to complete the closure. While the first term in  $h(\widehat{\cal P})$ was treated in the Ehrenfest case above, special attention is needed in dealing with the backreaction energy ${\cal B}=\int\big\langle i\hbar\{\widehat{\cal P},\widehat{H}\}\big\rangle\, \de^3q \de^3p$ in \eqref{KMod1}. A possible way to proceed is to factorize $\widehat{\cal P}={\cal F}\widehat{\sf P}$, where ${\cal F}=\operatorname{Tr}\widehat{\cal P}$ and $\operatorname{Tr}\widehat{\sf P}(\bq,\bp)=1$. Then, upon restricting to a quantum-classical Hamiltonian operator of the form $\widehat{H}=M^{-1}p^2/2+\widehat{\cal H}(q)$, we write
\[
\int\big\langle i\hbar\{\widehat{\cal P},\widehat{H}\}\big\rangle\, \de^3q \de^3p
=-{\operatorname{Tr}}\int \frac{i\hbar}2{\cal F}\big[\widehat{\sf P},\partial_\bp\widehat{\sf P}\big]\cdot\partial_\bq\widehat{\cal H}\, \de^3q \de^3p,
\]
where we have suitably projected onto the Hermitian part inside the trace. 

At this point, we can approximate  the classical distribution ${\cal F}$ by resorting to the \emph{cold-fluid ansatz}, that is ${\cal F}(\bq,\bp)=D(\bq)\delta(\bp-\bm(\bq)/D(\bq))$, where we omitted time dependence for compactness. We remark that this ansatz is adopted at this stage only as a modeling tool, while the thermodynamic internal energy $\mathscr{U}(D,s)$ will be accounted for in the total energy. {Thermal backreaction effects may also be incorporated by using more sophisticated warm-fluid ansatzes, likely increasing the level of complexity. As we strive to keep the latter as low as possible, in the present treatment thermal  effects will only be encoded within the solvent internal energy while they will be ignored in modeling the backreaction.}
With this approximation, we have
\[
{\operatorname{Tr}}\int\big\langle i\hbar\{\widehat{\cal P},\widehat{H}\}\big\rangle\, \de^3q \de^3p
\simeq-{\operatorname{Tr}}\int \frac{i\hbar}2D\partial_\bq\widehat{\cal H}\cdot\big[\widehat{\sf P},\partial_\bp\widehat{\sf P}\big]_{\bp=M\bu}\, \de^3q,
\]
where $\bu=M^{-1}\bm/D$  as in the previous section. 

The cold-fluid ansatz allows writing $D\widehat{\sf P}|_{\bp=M\bu}=\tilde\rho=D\hat\rho$, so that our closure problem amounts to finding a suitable closure for the quantity
\[
\widehat{\boldsymbol\kappa}=D\partial_\bp\widehat{\sf P}\big|_{\bp=M\bu}.
\]
In particular, we need to express the above quantity in terms of $\hat\rho$ and $\nabla\hat\rho$. In order to minimize the level of complexity, we will assume that $\widehat{\boldsymbol\kappa}$ depends linearly on $\nabla\hat\rho$. Furthermore, we also want to devise a closure that prevents the appearance of higher-order gradients when integrating by parts as ${\operatorname{Tr}}\int {i\hbar}[\hat\rho,\widehat{\kappa}^j]\partial_j\widehat{\cal H}/2\, \de^3q=-{\operatorname{Tr}}\int {i\hbar}\widehat{\cal H}([\partial_j\hat\rho,\widehat{\kappa}^j]+[\hat\rho,\partial_j\widehat{\kappa}^j])/2\, \de^3q$. Therefore, we are left with the option $\widehat{\boldsymbol\kappa}=-\boldsymbol{v}\times\nabla\hat\rho$, where the minus sign is inserted for later convenience and $\boldsymbol{v}(\bq,t)$ is added as an auxiliary variable. 

The dynamics of $\boldsymbol{v}$ represents the final step in the closure scheme. A simple proposal in \cite{GBTr-fluid} consists in writing $\boldsymbol{v}=c\nabla b$, so that 
\[
\widehat{\boldsymbol\kappa}=-c\nabla b\times\nabla\hat\rho
\]
and the final expression of the total energy becomes
\beq\label{HamHKoop}
{\sf h}(\bm,D,\tilde\rho,b,c,s)={\int}\left(\frac1{2M}\frac{m^2}{D}+D\mathscr{U}(D,s) + {\operatorname{Re}}{\operatorname{Tr}}\left(\tilde\rho\widehat{\cal H}+\frac{i\hbar}{D^2} c\tilde\rho\{\tilde\rho,\widehat{\cal H}\}_b\right)\right) \de^3q, 
\eeq
where the notation
\beq\label{Nambubkt}
\{A,B\}_b:=\nabla b\cdot\nabla A\times\nabla B
\eeq
identifies the Nambu Poisson bracket on physical space \cite{Nambu}. In this way, {given  the presence of the Poisson bracket $\{\cdot,\cdot\}_b$ in \eqref{HamHKoop}, we have ensured for the backreaction energy   to retain its original overall structure throughout the hydrodynamic closure}. Here,  $c$ and $b$ are scalar functions whose dynamics needs to be conveniently prescribed. In particular, the function $c(\bq,t)$ is crucial in ensuring convergence of the last integral term in the total energy. As such, it can be initialized as $c(\bq,0)=D(\bq,0)$ and we enforce hydrodynamic transport $\partial_t c+\bu\cdot\nabla c=0$ in such a way for $c$ to retain fast decay at infinity. {Then, the variable $c/D$ represents the Jacobian determinant of the Lagrangian flow \cite{GBTr-fluid}, thereby encoding information on its compressibility}. Likewise, considerations involving momentum balance and the transformation properties of the integrand in \eqref{HamHKoop} led in \cite{GBTr-fluid} to let $b$ evolve according to $\partial_t b+\bu\cdot\nabla b=0$. The variable $b$ is particularly crucial in determining the backreaction, which would vanish whenever $b$ is chosen as a constant. For this reason, $b$ was dubbed \emph{backreaction field} in \cite{GBTr-fluid}, where questions about its initial condition were also discussed. {A further physical characterization of the backreaction field and its accompanying variable $c$ requires additional work, possibly based on numerical experiments in three-dimensional space and analogies with advanced hydrodynamic models in fluid mechanics. As we will see, however, these extra variables $c$ and $b$ will be entirely omitted when dealing with polar solvation}. 
As we will see, these questions may be circumvented in the case of planar dynamics.

In the presence of two transported scalars $c$ and $b$, we can adopt the technique discussed in the previous section, thereby finalizing the closure by suitably extending the Poisson bracket \eqref{Ehr-brkt-adb}. Indeed, the latter becomes
\begin{align}\nonumber
\{f,k\}_\text{HK}=&{\int}\bm\cdot\left(\frac{\delta k}{\delta \bm}\cdot\nabla\frac{\delta f}{\delta \bm}-\frac{\delta f}{\delta \bm}\cdot\nabla\frac{\delta k}{\delta \bm}\right)\de ^3 q
\\\nonumber
&
-{\operatorname{Tr}}{\int} \tilde\rho\left( i\hbar^{-1}\left[\frac{\delta k}{\delta \tilde\rho},\frac{\delta f}{\delta \tilde\rho}\right]+\frac{\delta f}{\delta \bm}\cdot\nabla\frac{\delta k}{\delta \tilde\rho}-\frac{\delta k}{\delta \bm}\cdot\nabla\frac{\delta f}{\delta  \tilde\rho}\right)\de^3  q
\\\nonumber
&
-
{\int}D\left(\frac{\delta f}{\delta \bm}\cdot\nabla\frac{\delta k}{\delta D}-\frac{\delta k}{\delta \bm}\cdot\nabla\frac{\delta f}{\delta D}\right)\de ^3 q
-
\int \! s\operatorname{div}\!\left(\frac{\delta f}{\delta \bm}\frac{\delta k}{\delta s}-\frac{\delta k}{\delta \bm}\frac{\delta f}{\delta s}\right)\de ^3 q
\\
&
-
\int \! b\operatorname{div}\!\left(\frac{\delta f}{\delta \bm}\frac{\delta k}{\delta b}-\frac{\delta k}{\delta \bm}\frac{\delta f}{\delta b}\right)\de ^3 q
-
\int \! c\operatorname{div}\!\left(\frac{\delta f}{\delta \bm}\frac{\delta k}{\delta c}-\frac{\delta k}{\delta \bm}\frac{\delta f}{\delta c}\right)\de ^3 q,
\label{Koop-brkt-adb}
\end{align}
where the subscript `HK' stands for \emph{Hydrodynamic Koopman} model.
Here, the last two terms correspond to the hydrodynamic transport for the variables $b$ and $c$. These two terms are added to the last term in the third row, which by itself corresponds to entropy transport{, as discussed in \S\ref{sec:MQCHydroAdbEhr}}.
At this point, the relation $\dot{f}=\{f,{\sf h}\}_\text{HK}$ yields the generalized Hamiltonian equations for adiabatic Koopman hydrodynamics:
\begin{align}\label{HHam1adiK}
&\frac{\partial D}{\partial t}+\operatorname{div}\left(\frac{\delta \sf h}{\delta \bm}D\right)=0
\,,\qquad
\frac{\partial \tilde\rho}{\partial t}+\operatorname{div}\left(\frac{\delta \sf h}{\delta \bm}\tilde\rho\right)=-\frac{i}\hbar\left[\frac{\delta \sf h}{\delta \tilde\rho},\tilde\rho\right]
\,,\qquad
\frac{\partial s}{\partial t}+\frac{\delta \sf h}{\delta \bm}\cdot\nabla s=0,
\\\label{HHam2adiK}
&
\frac{\partial b}{\partial t}+\frac{\delta \sf h}{\delta \bm}\cdot\nabla b=0
\,,\qquad
\frac{\partial c}{\partial t}+\frac{\delta \sf h}{\delta \bm}\cdot\nabla c=0,
\\\label{HHam3adiK}
&\frac{\partial \bm}{\partial t}+\operatorname{div}\left(\frac{\delta \sf h}{\delta \bm}\bm\right)+\nabla\frac{\delta \sf h}{\delta \bm}\cdot\bm=\frac{\delta \sf h}{\delta s}\nabla s+\frac{\delta \sf h}{\delta b}\nabla b+\frac{\delta \sf h}{\delta c}\nabla c-D\nabla\frac{\delta \sf h}{\delta D}-{\operatorname{Tr}}\left(\tilde\rho\nabla\frac{\delta \sf h}{\delta \tilde\rho}\right),
\end{align}
where \eqref{HHam1adiK} are identical to \eqref{HHam1adi}-\eqref{HHam3adi}, while the momentum equation \eqref{HHam4adi} is now modified to retain the effects of the variables $b$ and $c$ obeying \eqref{HHam2adiK}. At this point, the final equations are obtained by expanding the functional derivatives. In doing so,  it is convenient to define 
\[
{\cal B}={\operatorname{Re}}{\operatorname{Tr}}{\int}\frac{i\hbar c}{D^2}\tilde\rho\{\tilde\rho,\widehat{\cal H}\}_b\,\de^3q={\operatorname{Re}}{\operatorname{Tr}}{\int}{i\hbar c}\hat\rho\{\hat\rho,\widehat{\cal H}\}_b\,\de^3q=:{\int}\varepsilon(c,\hat\rho,\nabla\hat\rho,\nabla b, \nabla\widehat{H})\,\de^3q,
\]
where $\hat\rho=\tilde\rho/D$. Then, upon introducing $\widehat{\bbeta}=\hbar c\hat\rho\nabla b$ {and using $\varepsilon={\operatorname{Re}}{\operatorname{Tr}}(i\,\widehat{\!\boldsymbol\beta} [\nabla{ \hat{\rho}},\times\nabla\widehat{ H}])/2$}, we rewrite
\begin{align*}
&\,\frac{\delta {\cal B}}{\delta b}\nabla b+\frac{\delta {\cal B}}{\delta c}\nabla c-{\operatorname{Tr}}\bigg(\tilde\rho\nabla\frac{\delta {\cal B}}{\delta \tilde\rho}\bigg)-D\nabla\frac{\delta {\cal B}}{\delta D}
\\
=
&\,
\frac{\partial \varepsilon}{\partial c}\nabla c-\nabla b\operatorname{div}\frac{\partial \varepsilon}{\partial \nabla b}+{\operatorname{Tr}}\bigg(\bigg(\frac{\partial \varepsilon}{\partial \hat\rho}
-\operatorname{div}\frac{\partial \varepsilon}{\partial \nabla\hat\rho}\bigg)\nabla\hat\rho\bigg)
\\
=
&\,
\nabla \varepsilon-\partial_j\bigg(\frac{\partial \varepsilon}{\partial \partial_j b}\nabla b+{\operatorname{Tr}}\bigg(\frac{\partial \varepsilon}{\partial \partial_j\hat\rho}\nabla\hat\rho\bigg)\bigg)
-
{\operatorname{Tr}}\bigg(\frac{\partial \varepsilon}{\partial \partial_j\widehat{H}}\partial_j\nabla\widehat{H}\bigg)
\\
=
&\,
\nabla \varepsilon-\partial_j{\operatorname{Tr}}\bigg(\frac{\partial \varepsilon}{\partial \widehat{\beta}_j}\widehat{\bbeta}+\frac{\partial \varepsilon}{\partial \partial_j\hat\rho}\nabla\hat\rho\bigg)
-
{\operatorname{Tr}}\bigg(\frac{\partial \varepsilon}{\partial \partial_j\widehat{H}}\partial_j\nabla\widehat{H}\bigg)
\\
=
&\,
{\operatorname{Tr}}\bigg(\frac{\partial \varepsilon}{\partial \widehat{\bbeta}}\times\operatorname{curl}\widehat{\bbeta}
-\nabla\hat\rho\operatorname{div}
\frac{\partial \varepsilon}{\partial \nabla\hat\rho}\bigg)
\\
=&\,
i{\operatorname{Re}}{\operatorname{Tr}}\big(
(\operatorname{curl}\widehat{\bbeta}\cdot\nabla\hat\rho) \nabla\widehat{H}
\big),
\end{align*}
where we used $\operatorname{div}({\partial \varepsilon}/{\partial \widehat{\bbeta}})=0$. Further evaluation of the functional derivatives of \eqref{HamHKoop} yields the system
\begin{equation}\label{QCeqs1}
\begin{aligned}
&MD\left(\frac{\partial}{\partial t}+\bu\cdot\nabla\right)\bu=-\nabla{\sf p}
-{\operatorname{Re}}{\operatorname{Tr}}\Big(\big(D\hat\rho+
{i\hbar} \operatorname{div}(c
    \hat{\rho}\nabla \hat{\rho}\times\nabla b)\big)\nabla  \widehat{\cal H}
\Big)
\\
&i\hbar D\left(\frac{\partial}{\partial t}+\bu\cdot\nabla\right)\hat\rho= \left[D\widehat{\cal H}+{i\hbar}\Big(c\{\hat\rho,\widehat{\cal H}\}_b+c\{\widehat{\cal H},\hat\rho\}_b-\frac{1}{2}\big[\{c, \widehat{\cal H}\}_b,\hat\rho\big]\Big),\hat\rho\right],
\\
&\left(\frac{\partial}{\partial t}+\bu\cdot\nabla\right)s=0
,\qquad\quad
\left(\frac{\partial}{\partial t}+\bu\cdot\nabla\right)b=0
,\qquad\quad
\left(\frac{\partial}{\partial t}+\bu\cdot\nabla\right)c=0,
\\
&\frac{\partial D}{\partial t} +\operatorname{div}(D\bu)=0\,.
\end{aligned}
\end{equation}

We observe that the system above involves only first-order gradients: no higher-order derivatives occur. {The right-hand side of the first equation involves two force terms beyond pressure: the first term inside the trace corresponds to the usual Hellmann-Feynman  force from Ehrenfest dynamics, while the second term involves the backreaction field $b$ and thus identifies backreaction forces. It is precisely this second backreaction term that restores quantum decoherence dynamics beyond the Ehrenfest model; this was recently shown in \cite{BaBeGBTr24,BeMaTr26} by a numerical implementation of the Koopman model in phase-space}.
Also, the second equation  indicates that the quantum state variable $\hat\rho$ evolves unitarily in the frame moving with the hydrodynamic flow, thereby preserving the positivity of $\hat\rho$ throughout the entire evolution. This property is shared with the Ehrenfest model from the previous section. Of particular interest are the Casimir invariants for the system \eqref{HHam1adiK}-\eqref{HHam3adiK} that is associated to the Poisson bracket \eqref{Koop-brkt-adb}. In addition to \eqref{Casimir1}, we have the following family of invariants: $
C_2=\int\!D\Gamma(b,c,s,\{\Lambda_n\})\,\de^3q$, where $\Lambda_n=(D^{-1}\nabla c\times\nabla b\cdot\nabla)^ns$ and $\Gamma$ is an arbitrary function of four variables. In turn, the hydrodynamic helicity \eqref{Helicity} is lost in the present case. In the  absence of entropy transport, one allows for the \emph{cross-helicity} invariant $\int c\nabla b\cdot\nabla \times(M\bu-\bA)\,\de^3q$, as noted in \cite{GBTr-fluid}.

A particular special case of the hydrodynamic system \eqref{QCeqs1} is given by its restriction to two-dimensional dynamics. This restriction can be obtained by eliminating the $z$-dependence on all variables except for $b$, which instead is written as $b(x,y,z)=\beta z$. In this case, upon denoting $\tilde{c}=\beta c$ the system \eqref{QCeqs1} becomes
\begin{equation}\label{QCeqs2}
\begin{aligned}
&MD\left(\frac{\partial}{\partial t}+\bu\cdot\nabla\right)\bu=-\nabla{\sf p}
-{\operatorname{Re}}{\operatorname{Tr}}\big((D\hat\rho+
{i\hbar} \{\tilde{c}
    \hat{\rho},{\hat{\rho}}\}_{xy})\nabla  \widehat{\cal H}
\big)
\\
&i\hbar D\left(\frac{\partial}{\partial t}+\bu\cdot\nabla\right)\hat\rho= \left[D\widehat{\cal H}+{i\hbar}\Big(\tilde{c}\{\hat\rho,\widehat{\cal H}\}_{xy}+\tilde{c}\{\widehat{\cal H},\hat\rho\}_{xy}-\frac{1}{2}\big[\{\tilde{c}, \widehat{\cal H}\}_{xy},\hat\rho\big]\Big),\hat\rho\right],
\\
&\frac{\partial D}{\partial t} +\operatorname{div}(D\bu)=0
,\qquad\quad
\left(\frac{\partial}{\partial t}+\bu\cdot\nabla\right)s=0
,\qquad\quad
\left(\frac{\partial}{\partial t}+\bu\cdot\nabla\right)\tilde{c}=0\,,
\end{aligned}
\end{equation}
where $\{A,B\}_{xy}=\partial_xA\partial_yB-\partial_xB\partial_yA$ is the canonical Poisson bracket on the plane. We observe that this planar restriction of the hydrodynamic system eliminates the backreaction field variable $b$, which instead becomes a constant $\beta$ that is absorbed by $\tilde{c}=\beta c$.



\subsection{Ideal solvation of a nitric oxide molecule in supercritical argon\label{sec:NOEx}}

Here, we specialize the equations above to study the evolution of a NO molecule in a nonpolar compressible Ar solvent. Our treatment extends the case of study in \cite{Egorov,HuEtAl12} to two spatial dimensions. In particular, the NO molecule is modeled as a two-level quantum system coupled to the classical  solvent. We also refer to \cite{RuHe88} for an analogous problem involving a bromine molecule solute Br$_2$.

\subsubsection{Model Hamiltonian and equations of motion}
In terms of the Pauli matrices, the Hamiltonian $\widehat{\cal H}$ is given as \cite{HuEtAl12}
\beq\label{NOArHam}
\widehat{\cal H}=\frac12(\phi_1+\phi_2)+\frac12(\phi_2-\phi_1)\widehat{\sigma}_3+C{*}(D-{D}_e),
\eeq
where the symbol $\ast$ denotes convolution and $D_e$ is a reference equilibrium profile. Also, upon denoting $q=|\bq|$,
\beq\label{ACF}
C(q)=k_BT\cos\bigg(\frac38q\bigg)\exp\bigg(-\frac{q}2\bigg),
\eeq
models the Ar \emph{autocorrelation function}, while the $\phi$ potentials deserve a separate discussion. In principle, one would use the Lennard-Jones and Buckingham-type potentials
\[
\phi_1(q)= 4\epsilon_{1} \bigg[ \left( \frac{\upsigma_{1}}{q} \right)^{12} - \left( \frac{\upsigma_{1}}{q} \right)^{6} \bigg]
,\qquad\quad
\phi_2(q)=\frac{\epsilon_{2}}{1 - 6/\alpha} \bigg[ \frac{6}{\alpha} \exp\left(\alpha\left(1 - q/r_{e}\right)\right) - \left( \frac{r_{e}}{q} \right)^{6} \bigg]
\]
with the numerical parameters given by Table I in \cite{HuEtAl12}. However, we observe that these potentials are particularly stiff and become singular at the origin. For this reason, these potentials are not used directly in numerical implementations and suitable regularizations  are instead preferred. In \cite{HuEtAl12}, a quadratic cutoff was performed in such a way to devise the regularized potential 
\[
\bar\phi_j(q)=(0.1-a_jq^2)\Theta(\bar{q}_{j}-r)+\phi_j(q)\Theta(r-\bar{q}_{j}),
\] 
where $\Theta$ is the Heaviside step function and $\bar{q}_j$ is the regularization cutoff radius for each $j=1,2$. Specific values of the regularization parameters were also given in \cite{HuEtAl12}.

Then, upon dropping the entropy dependence to restrict to isothermal fluid flows, the total energy \eqref{HamHKoop} is modified as
\begin{multline}\label{HamHKoop2}
{\sf h}(\bm,D,\tilde\rho,b,c)={\int}\left(\frac1{2M}\frac{m^2}{D}+D\mathscr{U}(D)+ \frac12D(\bar\phi_1+\bar\phi_2)+\frac12(D-{D}_e)C{\ast}(D-{D}_e)\right.
\\
+ \left.\frac12{\operatorname{Tr}}\Big(\widehat{\sigma}_3\tilde\rho(\bar\phi_2-\bar\phi_1)+\frac{i\hbar c}{D^2} \widehat{\sigma}_3\tilde\rho\big\{\tilde\rho,(\bar\phi_2-\bar\phi_1)\big\}_{xy}\Big)\right) \de^3q, 
\end{multline}
so that the Hamiltonian operator  is  replaced by $\widehat{\cal H}=(\bar\phi_1+\bar\phi_2)/2+(\bar\phi_2-\bar\phi_1)\widehat{\sigma}_3/2+C{*}(D-{D}_e)$. The final equations of motion for the isothermal variant of the model \eqref{QCeqs2} read
\begin{align}\nonumber
&M\Big(\frac{\partial}{\partial t}+\bu\cdot\nabla\Big)\bu=-\frac1D\nabla{\sf p}
-\frac{\bq}{2q}\bigg((\bar\phi_1'+\bar\phi_2')
+ 2C'{*}(D-D_e)
\\
&\hspace{8.25cm}
+(\bar\phi_2'-\bar\phi_1'){\operatorname{Tr}}\Big(\widehat{\sigma}_3\hat\rho+
\frac{i\hbar}D \widehat{\sigma}_3\{\tilde{c}
    \hat{\rho},{\hat{\rho}}\}_{xy}
\Big)  \bigg),
\label{NOArEqn1}
\\
&i\hbar \Big(\frac{\partial}{\partial t}+\bu\cdot\nabla\Big)\hat\rho= \frac12\left[(\bar\phi_2-\bar\phi_1)\widehat{\sigma}_3+\frac{i\hbar\tilde{c}}D(\bar\phi_2'-\bar\phi_1')\Big[\{\hat\rho,q\}_{xy}+\{{\ln}\sqrt{\tilde{c}},q\}_{xy}\hat\rho,\widehat{\sigma}_3\Big],\hat\rho\right],
\label{NOArEqn2}
\\
&\frac{\partial D}{\partial t} +\operatorname{div}(D\bu)=0
,\qquad\quad
\Big(\frac{\partial}{\partial t}+\bu\cdot\nabla\Big)\tilde{c}=0\,.
\label{NOArEqn3}
\end{align}
In this model, the barotropic pressure $\sf p$ with a polytropic index tuned to the monatomic bulk response  provides the thermodynamic anchor, while the non-local autocorrelation \eqref{ACF} accounts for the microscopic packing and excluded-volume effects.

We observe that the 2D Poisson bracket terms trigger the  rotational dynamics, thereby leading to full planar motion as long as the initial  conditions do not depend only on the radial coordinate $|\bq|$. Otherwise, one-dimensional radial solutions are preserved in time, in which case the $\tilde{c}$-equation decouples and the resulting 1D system reduces to a nonlocal variant of the Ehrenfest model. 

{
\subsubsection{Anisotropic pure-dephasing dynamics and  solvent response}

To explicitly capture the  breakdown of the Ehrenfest approximation, 
here we consider a planar solvation system in which dissipative effects are neglected for illustration purposes and we restore the angular dependence on the interatomic potentials. For example, one may replace $\bar\phi_j(q)\to\bar\phi_j(q)(1+\upepsilon\cos 2\theta)=\varphi_j(q,\theta)$, thereby augmenting the potential profiles with an explicit quadrupolar perturbation \cite{Thuis}. For completeness, we recall that in polar coordinates $\{A,B\}_{xy}=(\partial_qA\partial_\theta B-\partial_qB\partial_\theta A)/q$.

In order to shed some light on the dynamical process governed by equations \eqref{NOArEqn1}-\eqref{NOArEqn3} in the present configuration, let us start by focusing on the quantum dynamics  \eqref{NOArEqn2}. Recalling the second in \eqref{QCeqs2} and the expression \eqref{NOArHam} of the Hamiltonian $\widehat{\cal H}$, inserting the anisotropic potentials $\varphi_j(q,\theta)$ in \eqref{NOArEqn2} leads to
\[
 D\left(\frac{\partial}{\partial t}+\bu\cdot\nabla\right)\hat\rho= -\left[\frac{i}\hbar D(\varphi_2-\varphi_1)\widehat{\sigma}_3+\tilde{c}\{[\widehat{\sigma}_3,\hat\rho],V\}_{xy}+\frac{1}{2}\{\tilde{c}, \varphi_2-\varphi_1\}_{xy}\big[\widehat{\sigma}_3,\hat\rho\big],\hat\rho\right]
\]
so that, upon denoting $\langle\widehat{\sigma}_j\rangle=\operatorname{Tr}(\hat\rho\widehat{\sigma}_j)$ and using $\hat\rho=(\boldsymbol{1}+\langle\widehat{\bsigma}\rangle\cdot\widehat{\bsigma})/2$ to write
\[
\operatorname{ReTr}([\hat\rho,\widehat{\sigma}_j]\nabla[\widehat{\sigma}_3,\hat\rho])=-2\langle\widehat{\sigma}_3\rangle\nabla\langle\widehat{\sigma}_j\rangle
,\qquad\qquad
\operatorname{ReTr}([\hat\rho,\widehat{\sigma}_j][\widehat{\sigma}_3,\hat\rho])=2(\delta_{j3}-\langle\widehat{\sigma}_j\rangle\langle\widehat{\sigma}_3\rangle)
,
\]
we have
\begin{multline}\label{BVev}
D \left( \frac{\partial}{\partial t} + \bu \cdot \nabla \right) \langle\widehat{\bsigma}\rangle = \frac{2}{\hbar} D  (\varphi_2-\varphi_1)  \langle\widehat{\bsigma}\rangle \times \mathbf{e}_3  
\\
+ 2 \tilde{c} \langle\widehat{\sigma}_3\rangle  \{ \langle\widehat{\bsigma}\rangle, \varphi_2-\varphi_1 \}_{xy} + \{ \tilde{c}, \varphi_2-\varphi_1 \}_{xy} \langle\widehat{\bsigma}\rangle\times\big(\langle\widehat{\bsigma}\rangle\times\mathbf{e}_3\big)
.
\end{multline}
Here, $\mathbf{e}_3$ denotes the unit vertical. We observe that initial conditions $\langle\widehat{\sigma}_3\rangle|_{t=0}=\pm1$, corresponding to the ground and excited state, are preserved by the dynamics. Therefore, as usual in this case, the quantum dynamics becomes trivial and, according to the momentum equation 
\begin{multline}\label{NOArEqn2D}
M\Big(\frac{\partial}{\partial t}+\bu\cdot\nabla\Big)\bu=-\frac1D\nabla{\sf p}
-\frac12\bigg(\nabla(\varphi_1+\varphi_2)
+ 2\nabla C{*}(D-D_e)
\\
+\mathbf{e}_3\cdot\bigg(\langle\widehat{\bsigma}\rangle+
\frac{\hbar}{2D}\langle\widehat{\bsigma}\rangle{\times}\{\tilde{c},\langle\widehat{\bsigma}\rangle\}_{xy}
\bigg)  \nabla(\varphi_2-\varphi_1)\bigg),
\end{multline}
 the solvent evolves adiabatically under time-independent forces. To avoid this scenario, the authors in \cite{HuEtAl12} introduce a small electronic coupling parameter producing off-diagonal elements in the Hamiltonian $\mathcal{H}$. For the sake of model comparison, here we consider a different approach that does not change the form of the Hamiltonian operator and  instead resorts to the initial condition $\langle\widehat{\sigma}_3\rangle|_{t=0}=0$,  which yields  pure-dephasing evolution. Although this is hardly an  experimentally realizable initial state, it is convenient here to isolate  the solvent response that is induced by the quantum-classical backreaction forces. Since \eqref{BVev} leads to $D(\partial_t+{\bu\cdot\nabla})\langle\widehat{\sigma}_3\rangle=\{\varphi_2-\varphi_1,\tilde{c}(1-\langle\widehat{\sigma}_3\rangle^2)\}$, we observe that  the overall quantum expectation $\int D\langle\widehat{\sigma}_3\rangle q\de q\de \theta$ consistently remains  constant, while the  relation $\langle\widehat{\sigma}_3\rangle=0$ is no longer preserved by the dynamics \cite{GBTr-fluid}. Instead,  nonzero values of $\langle\widehat{\sigma}_3\rangle$ are developed over time which trigger time-evolving forces in \eqref{NOArEqn2D}. These forces drive the solvent to execute  anisotropic radial breathing within the localized density fields and this breathing motion is precisely the inertial response to the anisotropic solute-solvent interactions. 

While a systematic characterization of this  process is the subject of ongoing numerical work, here we demonstrate that the proposed framework significantly outperforms Ehrenfest dynamics, as the latter completely misses the local environmental response. Indeed, the Ehrenfest model equations in the present case read as
\begin{equation}\nonumber
\begin{aligned}
&M\Big(\frac{\partial}{\partial t}+\bu\cdot\nabla\Big)\bu=-\frac1D\nabla{\sf p}
-\frac12\bigg(\nabla(\varphi_1+\varphi_2)
+ 2\nabla C{*}(D-D_e)
+\langle\widehat{\sigma}_3\rangle
  \nabla(\varphi_2-\varphi_1)\bigg),
\\
&i\hbar \Big(\frac{\partial}{\partial t}+\bu\cdot\nabla\Big)\hat\rho= \frac12(\varphi_2-\varphi_1)\left[\widehat{\sigma}_3,\hat\rho\right],
\qquad\qquad
\frac{\partial D}{\partial t} +\operatorname{div}(D\bu)=0\,.
\end{aligned}
\end{equation}
In this case, the quantum evolution equation yields $(\partial_t+\bu\cdot\nabla)\langle\widehat{\sigma}_3\rangle=0$ so that the initial condition $\langle\widehat{\sigma}_3\rangle|_{t=0}=0$ is now preserved by the dynamics. In this case, the forces acting on the solvent are again time-independent and the dynamics is adiabatic. The absence of breathing motion in this configuration shows the crucial difference between the two models considered here: Ehrenfest dynamics fails to capture the time-dependent solvent response, which instead is retained by the model proposed here. Perhaps not surprisingly, we also notice that the initialization $\langle\widehat{\sigma}_3\rangle|_{t=0}=\pm1$ of the ground or excited state leads to exactly the same adiabatic solvent dynamics in the two models.

A comparison with fully quantum nonadiabatic hydrodynamics \cite{FoHoTr} may also be discussed and we refer to \cite{GBTr-fluid} for a general discussion on the corresponding pure-dephasing evolution. In that case, the initial conditions $\langle\widehat{\sigma}_3\rangle|_{t=0}=\pm1$ produce again  adiabatic solvent dynamics as above. Instead, the pure-dephasing dynamics generated by the initial condition $\langle\widehat{\sigma}_3\rangle|_{t=0}=0$ leads to time-dependent solute-solvent interactions, similarly to the model proposed here.

To conclude, we emphasize that, although here we have neglected dissipative effects  to expose the quantum-classical backreaction mechanism, restoring shear viscosity and Stokes drag into the momentum evolution \eqref{NOArEqn2D} does not alter the findings discussed above. Indeed, the appearance of the  breathing process in the solvent depends strictly on the underlying quantum evolution. Dissipative solvent effects are incorporated in later sections.
}


\section{Micropolar order  in ideal solvation hydrodynamics\label{sec:PolSolv}}

In the study of solvation phenomena, the solvent polarization plays a crucial role in realizing  solute-solvent coupling forces and statistical correlations. An attempt to retain polarization effects within the hydrodynamic approach was performed in \cite{BuBa06,ChBa91a}. In this case,  {the mixed quantum-classical approach is based on a hybrid density operator $\widehat{\cal P}=\widehat{\cal P}(\bq,\bp,\bmu,\bn)$ defined on the extended phase-space, which includes} the angular momentum $\bmu$ and orientation $\bn$ coordinates. In \cite{BuBa06}, the authors extended the hydrodynamic approach to consider the moments $\widehat{\boldsymbol{g}}_T=\int \bp \widehat{\cal P}\de^3p\de^3\mu$ and $\widehat{\boldsymbol{g}}_{\boldsymbol\omega}=\int \boldsymbol\mu \widehat{\cal P}\de^3p\de^3\mu$ identifying the translational and angular momentum density, respectively, on the extended space with coordinates $(\bq,\bn)$. The dependence on both these coordinates is necessary to retain statistical correlations beyond mean-field models and extensions thereof. However, retaining the dependence on the orientation coordinate throughout the entire set of hydrodynamic moments may lead to increased levels of computational complexity. Here, we will attempt to alleviate this complexity by making a further closure step so that the hydrodynamic variables are still defined on physical space while the quantum-classical density $\tilde{\rho}(\bq,\bn)=\int  \widehat{\cal P}\de^3p\de^3\mu$ retains the dependence on the orientation vector $\bn$. In particular, we will treat the hydrodynamic variables within the context of Eringen's \emph{micropolar fluid theory} \cite{Eringen}.

In formulating the closure model, one needs to extend the system \eqref{KMod1} to retain the dependence on the orientational degrees of freedom $(\bmu,\bn)$. This is readily done by extending the definitions of canonical Poisson bracket $\{\cdot,\cdot\}$ and Hamiltonian vector field $\bX_{\widehat{A}}$. This approach was developed extensively  in \cite{GBTr23} for the simpler case of quantum-classical spin systems, in the absence of orientation coordinate $\bn$. In  the present case we have
\beq\label{XVF}
\bX_{\widehat{A}}=\big(\partial_\bp\widehat{A},-\partial_\bq\widehat{A},-\bmu\times\partial_{\bmu}\widehat{A}-\bn\times\partial_{\bn}\widehat{A},-\bn\times\partial_{\bmu}\widehat{A}\big)
\qquad\text{and}\qquad
\{A,B\}=\boldsymbol{\nabla} A\cdot\bX_{B}
,
\eeq
where $\boldsymbol{\nabla} A=(\partial_\bq{A},\partial_\bp{A},\partial_{\bmu}{A},\partial_{\bn}{A})$ is the overall gradient on the extended phase space. With these new definitions, the system \eqref{KMod1} remains formally unchanged while the Hamiltonian operator is of the type $\widehat{H}=\widehat{H}(\bq,\bp,\bmu,\bn)$. At this point, one can proceed by following the steps from Section \ref{sec:MQCHydro1}. Once again, we will proceed gradually by first extending the Ehrenfest model to include polarization effects.

\subsection{Polar Ehrenfest hydrodynamics\label{sec:MicEhr}}

Extending the treatment in Section \ref{sec:MQCHydroEhr}, here we present the closure procedure leading to Ehrenfest hydrodynamics with classical micropolar order. As before, we neglect the backreaction energy term in the integral from \eqref{KMod1} and proceed by applying the chain rule. In this case, we introduce the following hydrodynamic moments:
\begin{eqnarray}\nonumber
D(\bq)={\operatorname{Tr}}{\int} \widehat{\cal P}\de^3p\de^3\mu{\de^3 n}
,\quad &&
\bm(\bq)={\operatorname{Tr}}{\int}\bp \widehat{\cal P}\de^3p\de^3\mu{\de^3 n}
,
\\\label{moms2}
L(\bq)={\operatorname{Tr}}{\int}\bmu \widehat{\cal P}\de^3p\de^3\mu{\de^3 n}
,&&
\tilde\rho(\bq,\bn)=\!{\int} \widehat{\cal P}\de^3p\de^3\mu,
\end{eqnarray}
so that the chain rule yields
\beq\label{funCR2}
\frac{\delta h}{\delta \widehat{\mathcal{P}}}=\frac{\delta \sf h}{\delta D}+\bp\cdot\frac{\delta \sf h}{\delta \bm}+\bmu\cdot\frac{\delta \sf h}{\delta \bL}+\frac{\delta \sf h}{\delta \tilde\rho}.
\eeq
Replacing the latter in the Hamiltonian equation  in \eqref{KMod1} and taking moments leads to the following hydrodynamic Hamiltonian equations:
\begin{align}\label{HHam1Pol}
&\frac{\partial \bm}{\partial t}+\operatorname{div}\bigg(\frac{\delta \sf h}{\delta \bm}\bm\bigg)+\nabla\frac{\delta \sf h}{\delta \bm}\cdot\bm=-D\nabla\frac{\delta \sf h}{\delta D}-\nabla\frac{\delta \sf h}{\delta \bL}\cdot\bL
-{\operatorname{Tr}}{\int}
\tilde\rho\nabla\frac{\delta \sf h}{\delta \tilde\rho}\,{\de^3 n},
\\&\label{HHam2Pol}
\frac{\partial \bL}{\partial t}+\operatorname{div}\bigg(\frac{\delta \sf h}{\delta \bm}\bL\bigg)=-\bL\times\frac{\delta \sf h}{\delta \bL}-{\operatorname{Tr}}{\int}\tilde\rho\bn\times\frac{\partial}{\partial \bn}\frac{\delta \sf h}{\delta \tilde\rho}\,{\de^3 n},
\\&\label{HHam3Pol}
i\hbar\frac{\partial \tilde\rho}{\partial t}+i\hbar\operatorname{div}\bigg(\tilde\rho\frac{\delta \sf h}{\delta \bm}\bigg)+{i\hbar}\frac{\partial}{\partial \bn}\cdot\left(\tilde\rho\frac{\delta \sf h}{\delta \bL}\times\bn\right)=\left[\frac{\delta \sf h}{\delta \tilde\rho},\tilde\rho\right],
\\&\label{HHam4Pol}
\frac{\partial D}{\partial t}+\operatorname{div}\left(\bu D\right)=0.
\end{align}
We emphasize the different notation between the spatial gradient $\nabla=\partial_\bq$ and the extended phase-space gradient $\boldsymbol{\nabla}$ introduced {in \eqref{XVF}}. 
From the third term in \eqref{HHam3Pol}, we observe that the orientational evolution of $\tilde\rho$ is purely rotational. Consequently, 
equation \eqref{HHam3Pol} ensures solutions of the type $\tilde\rho(\bq,\bn,t)=\tilde{\varrho}(\bq,\bn,t)\delta(\|\bn\|-1)$, so that  $\bn$  effectively plays the role of  a unit vector.

In the presence of polarization, we can consider the following form of quantum-classical Hamiltonian operator:
\beq\label{MPHam}
\widehat{H}=\frac{p^2}{2M}+\frac{\mu^2}{2J}+\widehat{\cal H}(\bq,\bn)
\eeq
where $J$ is a {rotational} microinertia constant that is treated here as a simple numerical parameter \cite{BuBa06}.
Then, the first term in the total energy $h(\widehat{\mathcal{P}})$ from \eqref{KMod1} may be expressed in terms of the hydrodynamic moments by simply extending the procedure in \eqref{ClosHamEhr1}, that is
\begin{multline*}
{\operatorname{Tr}}{\int}\left(\frac{p^2}{2M}+\frac{\mu^2}{2J}+\widehat{\cal H}\right)\widehat{\cal P} \de^3q \de^3p\de^3\mu{\de^3 n}
\\
\simeq{\int}\left(\frac1{2M}\frac{m^2}{D}+\frac1{2J}\frac{L^2}{D}+D\mathscr{E}(D,\bP) + {\operatorname{Tr}}{\int}\tilde\rho\widehat{\cal H}{\de^3 n}\right) \de^3q={\sf h},
\end{multline*}
where ${\sf h}={\sf h}(\bm,D,\bL,\tilde\rho)$ is the total energy functional expressed in terms of the hydrodynamic variables. At this stage, since we are dealing with Ehrenfest dynamics, we  neglect backreaction energy contributions. 
Notice that in this case we have inserted the classical \emph{free energy} $\mathscr{E}$, which for polarized fluids depends also on $\tilde\rho$ via the polarization density 
\[
\bP=\alpha{\operatorname{Tr}}{\int} \bn\tilde\rho{\de^3 n},
\]
where $\alpha$ is the present notation for the electric dipole moment \cite{BaJa10,BuBa06}.
In practice, the free energy $\mathscr{E}$ retains nonlinear and possibly nonlocal terms associated to the dipole-dipole correlations within the classical solvent. 
To proceed further within the Ehrenfest model, we  evaluate functional derivatives so that the equations of motion in terms of  the variables \eqref{urhovars}  read
\begin{align}\label{EhrHam1Pol}
&\frac{\partial \bu}{\partial t}+\bu\cdot\nabla\bu=-\frac1{MD}\nabla{\sf p}
-
\frac1{MD}\nabla\left(D\frac{\partial \mathscr{E}}{\partial \bP}\right)\cdot\bP
-\frac1{M}{\operatorname{Tr}}{\int}
\hat\rho\nabla\widehat{\cal H}\,{\de^3 n},
\\&\label{EhrHam2Pol}
\frac{\partial \bL}{\partial t}+\operatorname{div}(\bu\bL)=-D\bP\times\frac{\partial \mathscr{E}}{\partial \bP}-D{\operatorname{Tr}}{\int}\hat\rho\bn\times\frac{\partial \widehat{\cal H}}{\partial \bn}\,{\de^3 n},
\\&\label{EhrHam3Pol}
i\hbar\frac{\partial \hat\rho}{\partial t}+i\hbar\bu\cdot\nabla\hat\rho+\frac{i\hbar}{JD}\bL\cdot\bn\times\frac{\partial\hat\rho}{\partial \bn}=\big[\widehat{\cal H},\hat\rho\big],
\\&\label{EhrHam4Pol}
\frac{\partial D}{\partial t}+\operatorname{div}\left(\bu D\right)=0,
\end{align}
where ${\sf p}=D^2\partial\mathscr{E}/\partial D$ and we have used the chain rule to write $\delta {\sf h}/\delta\tilde\rho=\widehat{\cal H}+\alpha D\bn\cdot\partial \mathscr{E}/\partial \bP$. From equation \eqref{EhrHam3Pol}, it follows that $(\partial_t+\bu\cdot\nabla)(\bP/D)=J^{-1}D^{-1}\bL\times\bP/D$ and an analogous result holds in the general case of equation \eqref{HHam3Pol}. In this way, if $|\bP|/D=\alpha$ initially, then the same holds at all times. 
In the absence of a quantum subsystem, one has $\widehat{\cal H}={\cal H}\hat{\boldsymbol{1}}$ and the state variable $\hat\rho$ is replaced by its trace $\tilde{D}(\bq,\bn)=\operatorname{Tr}\hat\rho(\bq,\bn)$. Then, the resulting system describes an ideal polarized fluid with microscopic multipolar order \cite{Tronci}.

As in Section \ref{sec:MQCHydroEhr}, the new Poisson structure $\{\cdot,\cdot\}_\text{H-MQCP}$ underlying the system \eqref{EhrHam1Pol}-\eqref{EhrHam4Pol} is revealed explicitly upon evaluating the time derivative $\de{f}/\de t={\int}(\delta f/\delta D)\partial_t D\de^3 q+{\int}(\delta f/\delta\bm)\cdot\partial_t \bm\de^3 q+{\int}(\delta f/\delta\bL)\cdot\partial_t \bL\de^3 q+{\operatorname{Tr}}{\int}(\delta f/\delta\tilde\rho)\partial_t \tilde\rho\de^3q\de^3 n$ of an arbitrary functional  along the dynamics prescribed by \eqref{HHam1Pol}-\eqref{HHam4Pol}. Using $\dot{f}=\{f,h\}_\text{H-MQCP}$, one obtains
\begin{multline}
\{f,k\}_\text{H-MQCP}=\!{\int}\bm\cdot\left(\frac{\delta k}{\delta \bm}\cdot\nabla\frac{\delta f}{\delta \bm}-\frac{\delta f}{\delta \bm}\cdot\nabla\frac{\delta k}{\delta \bm}\right)\de ^3 q
-
{\int}D\left(\frac{\delta f}{\delta \bm}\cdot\nabla\frac{\delta k}{\delta D}-\frac{\delta k}{\delta \bm}\cdot\nabla\frac{\delta f}{\delta D}\right)\de ^3 q
\\
-{\operatorname{Tr}}{\int} \tilde\rho\left( \frac{i}\hbar\left[\frac{\delta k}{\delta \tilde\rho},\frac{\delta f}{\delta \tilde\rho}\right]
+\frac{\delta f}{\delta \bm}\cdot\nabla\frac{\delta k}{\delta \tilde\rho}
-\frac{\delta k}{\delta \bm}\cdot\nabla\frac{\delta f}{\delta  \tilde\rho}
-\bn\cdot\bigg(\frac{\delta f}{\delta \bL}\times\frac{\partial}{\partial \bn}\frac{\delta k}{\delta \tilde\rho}
-
\frac{\delta k}{\delta \bL}\times\frac{\partial}{\partial \bn}\frac{\delta f}{\delta \tilde\rho}\bigg)
\right)\de^3  q\de^2  n
\\
-
{\int} \bL\cdot\left( \frac{\delta k}{\delta \bL}\times\frac{\delta f}{\delta \bL}
+\frac{\delta f}{\delta \bm}\cdot\nabla\frac{\delta k}{\delta \bL}
-\frac{\delta k}{\delta \bm}\cdot\nabla\frac{\delta f}{\delta  \bL}
\right)\de^3  q.
\label{Ehr-brktPol}
\end{multline}
To avoid a further increase of the level of complexity, here we do not consider adding the entropy transport, which can be easily included by following the procedure from Section \ref{sec:MQCHydroAdbEhr}. A direct verification shows that, upon writing $\tilde\rho(\bq,n)=\Upsilon(\bq,n)\Upsilon(\bq,n)^\dagger$, the Poisson bracket \eqref{Ehr-brktPol} allows the following family of Casimir invariants:
\[
C_3={\int} D\,\Xi(|\bS|)\de^3x
,\qquad\text{where}\qquad
\bS=\frac1D\left({\bL+{\hbar\operatorname{Im}}}{\int}\Upsilon^\dagger(\bn\times\partial_\bn)\Upsilon\,{\de^3 n}\right)
\]
and $\Xi$ is any function of one variable. Alternatively, {as discussed at the end of \S\ref{sec:MQCHydroEhr}}, one can  resort again to the Uhlmann representation \cite{Bondar,Tronci19,Uhlmann} by writing $\tilde\rho(\bq,n)=\mathcal{W}(\bq,n)\mathcal{W}(\bq,n)^\dagger$, so that in that case $\bS=D^{-1}\big(\bL+\hbar\operatorname{ImTr}{\int}\mathcal{W}^\dagger(\bn\times\partial_\bn)\mathcal{W}\,{\de^3 n}\big)$.

\subsection{Polarized hydrodynamic closure of the Koopman model\label{sec:hybPolKoop}}

While the Ehrenfest model with micropolar order captures essential features of quantum-classical coupling in polarized hydrodynamics, further work is needed in order to retain backreaction effects and their resulting decoherence associated to quantum-classical correlations. To this purpose, we will now restore the backreaction energy  given by the second term in the energy integral from \eqref{KMod1}. In the present setting, with the notation as in \eqref{XVF}, the backreaction integral reads
\begin{align}\nonumber
{\int}\big\langle i\hbar\{\widehat{\cal P},\widehat{H}\}\big\rangle \de^3q \de^3p\de^3\mu{\de^3n}
&=
{\operatorname{Tr}}{\int} \frac{i\hbar}2{\cal F}[\widehat{\sf P},\boldsymbol\nabla\widehat{\sf P}]\cdot\bX_{\widehat{H}} \de^3q \de^3p\de^3\mu{\de^3n}
\\\label{BRNTranPol}
&=
{\operatorname{Tr}}{\int} \frac{i\hbar}2{\cal F}\left(\partial_\bq{\widehat{\cal H}}\cdot[\partial_\bp\widehat{\sf P} ,\widehat{\sf P}]
+
\bn\cdot[\widehat{\sf P},\partial_{\bmu}\widehat{\sf P}]\times\partial_{\bn}\widehat{\cal H}
\right)
\de^3q \de^3p\de^3\mu{\de^3n}
\end{align}
As {done} before, we have used \eqref{MPHam} and factorized  $\widehat{\cal P}={\cal F}\widehat{\sf P}$, with ${\cal F}=\operatorname{Tr}\widehat{\cal P}$ and $\operatorname{Tr}\widehat{\sf P}(\bq,\bp)=1$. Before proceeding further,  we will make a further approximation to alleviate the level of complexity {that results}  from the combination of backreaction terms associated to both translational and orientational degrees of freedom. {These  degrees of freedom are associated  to the first and the second term  in the parenthesis of \eqref{BRNTranPol}, respectively. We} recall that in the theory of polar solvation the solute-solvent correlations are mostly due to  polarization effects, while translational degrees of freedom play a much less important role. Therefore, we are naturally led to neglect the backreaction associated to translational degrees of freedom while retaining instead the terms corresponding to the orientation coordinate. {Notice that, while this approximation neglects certain energy contributions, it does not affect the overall Hamiltonian structure of the resulting model: this is due to the fact that both models share exactly the same Poisson bracket \eqref{Ehr-brktPol}. As we will see below, an important advantage of this approximation is that the backreaction field $b$ from \S\ref{sec:KoopHyd1} and its accompanying variable $c$ are both eliminated from the model. We refer to  Appendix \ref{app:translbackr} for a more general treatment beyond this approximation}. Here, neglecting translational backreaction  amounts to discarding the first term in the parenthesis of \eqref{BRNTranPol}, that is 
\beq\label{notranslapprox}
{\operatorname{Tr}}{\int} {i\hbar}{\cal F}\partial_\bq{\widehat{\cal H}}\cdot[\partial_\bp\widehat{\sf P} ,\widehat{\sf P}]
\de^3q \de^3p\de^3\mu{\de^3n}\simeq0.
\eeq

To proceed further, once again we approximate  the classical distribution ${\cal F}$ by the {cold-fluid ansatz}, which in this case reads ${\cal F}(\bq,\bp,\bmu,\bn)=\tilde{D}\,\delta(\bp-\bm/ D)\delta(\bmu-\bL/ D)$, where 
\[
\tilde{D}(\bq,\bn)={\operatorname{Tr}} \tilde\rho(\bq,\bn),
\] 
so that $D(\bq)=\int \tilde{D}(\bq,\bn){\de^3 n}$. Therefore, we evaluate
\[
{\operatorname{Tr}}{\int} {i\hbar} {\cal F}\bn\cdot[\widehat{\sf P},\partial_{\bmu}\widehat{\sf P}]\times\partial_{\bn}\widehat{\cal H} 
\de^3q \de^3p\de^3\mu{\de^3n}
=
{\operatorname{Tr}}{\int} {i\hbar} \tilde{D}\partial_{\bn}\widehat{\cal H} \cdot [\widehat{\sf P},\bn\times\partial_{\bmu}\widehat{\sf P}]|_{(\bp,\bmu)=(\bm,\bL)/D}\,
\de^3q {\de^3n},
\]
and the closure is completed by expressing the quantity $\bn\times\partial_{\bmu}\widehat{\sf P}|_{(\bp,\bmu)=(\bm,\bL)/D}$ in terms of the hydrodynamic variables in \eqref{moms2}. Following the treatment in Section \ref{sec:KoopHyd1},  we assume linearity  in $\partial_\bn(\tilde\rho/D)$ and avoid the appearance of higher-order derivatives. In particular, we write 
\[
\bn\times\frac{\partial\widehat{\sf P}}{\partial \bmu}\bigg|_{(\bp,\bmu)=(\bm,\bL)/D\!}\simeq\bn\times\frac{\partial}{\partial \bn}\frac{\tilde\rho}{\tilde{D}},
\]
so that
\[
{\operatorname{Tr}}{\int} {i\hbar} {\cal F}\bn\cdot[\widehat{\sf P},\partial_{\bmu}\widehat{\sf P}]\times\partial_{\bn}\widehat{\cal H} 
\de^3q \de^3p\de^3\mu{\de^3n}
\simeq
{\operatorname{Tr}}{\int} \frac{i\hbar}{\tilde{D}}\,\bn\cdot[\tilde\rho,\partial_\bn\tilde\rho]\times\partial_\bn\widehat{\cal H}\, \de^3q {\de^3n}.
\]
We observe that, unlike the nonpolar treatment discussed in Section \ref{sec:KoopHyd1}, here no auxiliary fields have been introduced so that  the number of dynamical variables in this closure is the same as in polar Ehrenfest hydrodynamics.
In conclusion, we have the closure
\beq\label{KoopClosure}
{\int}\big\langle i\hbar\{\widehat{\cal P},\widehat{H}\}\big\rangle \de^3q \de^3p\de^3\mu{\de^3n}
\simeq{\operatorname{Re}\operatorname{Tr}}
{\int}  \frac{i\hbar}{\tilde{D}}\tilde\rho\{\tilde\rho,\widehat{\cal H}\}_{S^2\,}
\de^3q {\de^3n},
\eeq
where $\{\cdot,\cdot\}_{S^2}$ denotes the Poisson bracket on the sphere, that is $\{A,B\}_{S^2}=\bn\cdot\partial_\bn A\times\partial_\bn B$. In this way, the correlations associated to the translational degrees of freedom are treated exactly as in the Ehrenfest model from Section \ref{sec:MicEhr}. Instead, the backreaction  term ${\operatorname{ReTr}}
\int  {i\hbar}{\tilde{D}}^{-1}\tilde\rho
\{\tilde\rho,\widehat{\cal H}\}_{S^2\,}
\de^3q {\de^3n}$ in the total energy
\[
{\sf h}={\int}\left(\frac1{2M}\frac{m^2}{D}+\frac1{2J}\frac{L^2}{D}+D\mathscr{E}(D,\bP) + {\operatorname{ReTr}}{\int}\tilde\rho\bigg(\widehat{\cal H}+\frac{i\hbar}{\tilde{D}}\{\tilde\rho,\widehat{\cal H}\}_{S^2}\bigg){\de^3 n}\right) \de^3q
\] 
produces the forces and torques  that are responsible for those solute-solvent correlations that are associated to the orientational dynamics. In this sense, the resulting equations of motion comprise a \emph{hybrid Ehrenfest-Koopman} model of polar solvation.

In this hybrid  model, the set of variables is the same as in the Ehrenfest model and the equations of motion are written using appropriate rearrangements as follows. Upon  denoting 
\[
\mathscr{D}_\bn=\bn\times\partial_\bn,
\] 
introducing $\hat\varrho=\tilde\rho/\tilde{D}$ and 
\begin{align*}
{\cal B}_{2}=&\, {\operatorname{ReTr}}
{\int}  \frac{i\hbar}{\tilde{D}}\tilde\rho\{\tilde\rho,\widehat{\cal H}\}_{S^2\,}
\de^3q {\de^3n}
\\
=&\, {\operatorname{ReTr}}
{\int} {i\hbar}{\tilde{D}}\hat\varrho\{\hat\varrho,\widehat{\cal H}\}_{S^2\,}
\de^3q {\de^3n}
\\
=&\!: {\int }\varepsilon_2(\tilde{D},\hat\varrho,\mathscr{D}_\bn\hat\varrho,\partial_\bn\widehat{\cal H})\de^3q {\de^3n},
\end{align*}
we have
\begin{align*}
&-{\operatorname{Tr}}
{\int}\tilde\rho\nabla\frac{\delta {\cal B}_2}{\delta \tilde\varrho}{\de^3n}
-{\int}\tilde{D}\nabla\frac{\delta {\cal B}_2}{\delta \tilde{D}}{\de^3n}
\\
=&\
{\operatorname{Tr}}
{\int}\left(\frac{\partial \varepsilon_2}{\partial \hat\varrho}-\mathscr{D}_\bn\cdot\frac{\partial \varepsilon_2}{\partial \mathscr{D}_\bn\hat\varrho}\right)\nabla\hat\varrho{\de^3n}
-\nabla{\int}\tilde{D}\frac{\partial\varepsilon_2}{\partial \tilde{D}}{\de^3n}
+\int\frac{\partial\varepsilon_2}{\partial \tilde{D}}\nabla\tilde{D}{\de^3n}
\\
=&\
\nabla{\int}\bigg(\varepsilon_2-\tilde{D}\frac{\partial\varepsilon_2}{\partial \tilde{D}}\bigg){\de^3n}
+
{\operatorname{Tr}}
{\int}\bigg(\partial_\bn\cdot\frac{\partial \varepsilon_2}{\partial \partial_\bn\widehat{\cal H}}\bigg)\nabla\widehat{\cal H}{\de^3n}
\\
=&\
-
{\operatorname{ReTr}}
{\int}i\hbar\{\tilde{D}\hat\varrho,\hat\varrho\}_{S^2}\nabla\widehat{\cal H}{\de^3n}.
\end{align*}
By proceeding analogously, we also find
\begin{align*}
-{\operatorname{Tr}}
{\int}\tilde\rho\mathscr{D}_\bn\frac{\delta {\cal B}_2}{\delta \tilde\rho}{\de^3n}
-{\int}\tilde{D}\mathscr{D}_\bn\frac{\delta {\cal B}_2}{\delta \tilde{D}}{\de^3n}
&=
-
{\operatorname{ReTr}}
{\int}i\hbar\{\tilde{D}\hat\varrho,\hat\varrho\}_{S^2}\mathscr{D}_\bn\widehat{\cal H}{\de^3n}.
\end{align*}
Then,  defining the quantum-classical pseudo-density operator 
\beq\label{pseudoD}
\widehat{\cal D}=\tilde\rho+\frac{i\hbar}2\bigg(\bigg\{\tilde\rho,\frac{\tilde\rho}{\tilde{D}}\bigg\}_{\!S^2}-\bigg\{\frac{\tilde\rho}{\tilde{D}},\tilde\rho\bigg\}_{\!S^2}\bigg)
\eeq
for compactness of notation, the hybrid Ehrenfest-Koopman model of polar solvation reads
\begin{align}\label{EhrKoopHam1Pol}
&MD\left(\frac{\partial}{\partial t}+\bu\cdot\nabla\right)\bu=-\nabla{\sf p}-\nabla\left(D\frac{\partial \mathscr{E}}{\partial \bP}\right)\cdot\bP
-{\operatorname{Tr}}{\int}\widehat{\cal D}\nabla  \widehat{\cal H}\,
{\de^3 n}
\\\label{EhrKoopHam2Pol}
&\,\frac{\partial \bL}{\partial t}+\operatorname{div}(\bu\bL)=-D\bP\times\frac{\partial \mathscr{E}}{\partial \bP}
-{\operatorname{Tr}}{\int}\widehat{\cal D}\mathscr{D}_\bn \widehat{\cal H}\,{\de^3 n}
\\
&\,i\hbar\left(\frac{\partial\tilde\rho}{\partial t}+{\operatorname{div}}(\bu\tilde\rho)+\frac{\bL}{JD}\cdot\mathscr{D}_\bn\tilde\rho\right)
=
[\widehat{\cal H},\tilde\rho]
+\frac1{\tilde{D}}\left[
 \{\tilde\rho,\widehat{\cal H}\}_{S^2}+\{\widehat{\cal H},\tilde\rho\}_{S^2}+\frac{1}{2}\big[\{\tilde{D}, \widehat{\cal H}\}_{S^2},\tilde\rho\big]
,\tilde\rho\right]
\\&\label{EhrKoopHam4Pol}
\frac{\partial D}{\partial t}+\operatorname{div}\left(\bu D\right)=0,
\end{align}
where we recall that ${\sf p}=D^2\partial\mathscr{E}/\partial D$,  $\mathscr{D}_\bn=\bn\times\partial_\bn$, and $\tilde{D}=\operatorname{Tr}\tilde\rho$.  Notice that, unlike the Ehrenfest case, here it is more convenient to express the equations in terms of the full density operator $\tilde\rho$ rather than $\hat\rho=\tilde\rho/D$. The Hamiltonian structure of \eqref{EhrKoopHam1Pol}-\eqref{EhrKoopHam4Pol}  is the same as in \eqref{Ehr-brktPol}.

{
Before specializing the above set of equations to the  case of  dielectric solvation, a few considerations are in place about their computational complexity. It is clear that the appearance of the pseudo-density operator \eqref{pseudoD} in the model raises the complexity well beyond Ehrenfest dynamics. This should not come as a surprise: as the model captures more and more phenomenology due to finer scales, the latter  naturally lead to a higher computational complexity. A raise in complexity beyond Ehrenfest is also found in the model \eqref{burghmod} since the latter involves a higher number of field variables. In that case, the level of complexity depends on the closure adopted for the stress tensor $\widehat\Pi$. Alternatively, in the model \eqref{EhrKoopHam1Pol}-\eqref{EhrKoopHam4Pol}, the complexity associated to the pseudo-density operator \eqref{pseudoD} is accompanied by a number of variables that is the same as in the Ehrenfest model. In addition to this lower number of variables, it is worth to mention that the latter are  defined on physical space with the only exception of the quantum state density $\tilde\rho(\bq,\bn,t)$, which also involves the orientational coordinate $\bn$. One may also follow similar steps to those in \cite{BuBa06,ChBa91} and proceed to a further level of reduction by replacing the quantum density variable by the \emph{quantum-classical polarization density $\widehat{\bf P}(\bq,t)=\alpha\int \bn\tilde\rho(\bq,\bn,t)\de^3n$}, in such a way that the entire set of variables is taken onto physical space. Even without this further step, the system \eqref{EhrKoopHam1Pol}-\eqref{EhrKoopHam4Pol} can be solved on a lower dimensional grid than that needed  for  the model  in \cite{BuBa06}, which involves the operator-valued momentum $\widehat{\boldsymbol{g}}_T(\bq,\bn,t)$ and angular momentum   $\widehat{\boldsymbol{g}}_{\boldsymbol\omega}(\bq,\bn,t)$, thereby indicating a  computational advantage in the approach presented here. A thorough investigation of these aspects, however, requires a computational comparison study which is beyond the purpose of this work.
}


\subsection{Nonlocal dielectric solvation and  Marcus hydrodynamics\label{MarcusHyd}}
Having illustrated the general setting and provided the equations of motion for a general Hamiltonian functional including backreaction, we now proceed to consider the energy associated to polar solvation. In this context, the  electrostatic interactions of the solvent are associated to the Poisson equation $\epsilon_0{\Delta}\Phi=\operatorname{div}\bP$, so that ${\mathbf{E}=-\epsilon_0^{-1}\nabla\operatorname{div}\Delta^{-1}\bP}$. Also, the quantum-classical Hamiltonian operator is written as ${\widehat{\cal H}=\widehat{H}_Q-\bn\cdot\widehat{\bf E}}$, where $\widehat{H}_Q$ is a purely quantum operator independent of spatial coordinates and we have defined ${\widehat{\bf E}=-\epsilon_0^{-1}\nabla\operatorname{div}\Delta^{-1}(\zeta\widehat{\bmu})}$. Here, $\zeta(\bq)$ is the fixed solute density,  $\widehat{\bmu}$ is the solute dipole operator array, and $\epsilon_0$ is the vacuum dielectric constant. With this notation, we write the total energy as \cite{BuBa06,KiHy90,KiHy90b}:
\begin{multline}
{\sf h}={\int}\bigg(\frac1{2}\frac{m^2}{MD}+\frac1{2}\frac{L^2}{JD}+D\mathscr{U}(D) +{\frac12(D-{D}_e)C{\ast}(D-{D}_e)}
+ \frac{|\bP|^2}{2\chi}-\frac1{2\epsilon_0}(\operatorname{div}\bP)\Delta^{-1}(\operatorname{div}\bP)
\\+{\operatorname{Tr}}(\tilde\rho\widehat{H}_Q)
+\frac{\alpha}{\epsilon_0}{\operatorname{ReTr}}{\int}\tilde\rho\bigg(\bn\cdot\nabla\operatorname{div}\Delta^{-1}(\zeta\widehat{\bmu})
+
\frac{i\hbar}{\tilde{D}}\bigg\{\tilde\rho,\bn\cdot\nabla\operatorname{div}\Delta^{-1}(\zeta\widehat{\bmu})\bigg\}_{S^2}\bigg){\de^3n}\bigg)
 \de^3q
\end{multline}
Here, $C(\bq)$ identifies the autocorrelation function for the solvent, similarly to the treatment in Section \ref{sec:NOEx}. Moreover, $\chi$ is the intrinsic orientational susceptibility of the solvent so that the fifth term in the first line is the solvent reorganization potential. The sixth term represents the solvent Coulomb energy associated to the solvent electric field ${\mathbf{E}=-\epsilon_0^{-1}\nabla\operatorname{div}\Delta^{-1}\bP}$. 
We remark that the solute density $\zeta(\bq)$ is typically taken as a delta function to model a point-like  molecule, although here we will avoid the appearance of singular forces by considering a smooth shape function such as a Gaussian centered at the solute location.

In this case, the solvent fluid momentum equation \eqref{EhrKoopHam1Pol}
\rem{ 
of motion must be found using the general Hamiltonian form \eqref{HHam1Pol}-\eqref{HHam4Pol}, upon evaluating the functional derivatives of the total energy. Upon introducing the symmetric operator $\Bbb{T}_{jk}=\partial_j\partial_k\Delta^{-1}$, so that $\widehat{\cal H}=D\tilde{D}^{-1}\widehat{\mu}_jn_k\Bbb{T}_{kj}\tilde{D}/\epsilon_0$,  equation \eqref{HHam1Pol} leads to
\begin{multline*}
MD\left(\frac{\partial}{\partial t}+\bu\cdot\nabla\right)\bu=-\nabla {\sf p}-\nabla\frac{\delta \sf h}{\delta \bP}\cdot\bP
-{\operatorname{Tr}}{\int}\widehat{\cal D}(\nabla  \widehat{\cal H}-\partial_{\tilde{D}}\widehat{\cal H}\nabla\tilde{D}-\partial_{{D}}\widehat{\cal H}\nabla{D})\,
{\color{RedOrange}\de^3 n}
\\
-\nabla{\operatorname{Tr}}{\int} \widehat{\cal D}\big(\tilde{D}\partial_{\tilde{D}}\widehat{\cal H}+D\partial_{{D}}\widehat{\cal H}\big){\color{RedOrange}\de^3 n}
-{\int}\tilde{D}\nabla {\operatorname{Tr}}\big(\Bbb{T}:\widehat{\mathscr{D}}\partial_{\Bbb{T}\tilde{D}}\widehat{\cal H}\big){\color{RedOrange}\de^3 n}
\end{multline*}
so that evaluating derivatives and using $\operatorname{curl}\bE=\operatorname{curl}\widetilde{\bE}=0$ yields
} 
reads
\beq
MD\bigg(\frac{\partial}{\partial t}+\bu\cdot\nabla\bigg)\bu=-\nabla\bigg({\sf p}+\frac{|\bP|^2}{2\chi}\bigg)-{D}\nabla C{*}(D-D_e)+\bP\cdot\nabla\bE
+\alpha
{\operatorname{Tr}}{\int}\widehat{\cal D}\bn\cdot\nabla\widehat{\bE}
{\de^3 n},
\eeq
where the {third}  term on the right-hand side is the \emph{Kelvin polarization force} generated by the inhomogeneity of the solvent's own electric field, while the {fourth} term is a \emph{dielectrophoretic force} that is excerpted on the dielectric solvent by the non-uniform electric field generated by the solute dipole. 
In addition, the  equations  \eqref{EhrKoopHam2Pol}-\eqref{EhrKoopHam4Pol} become
\begin{align}\label{MarcusHyd1PolA}
&\,\frac{\partial \bL}{\partial t}+\operatorname{div}(\bu\bL)=\bP\times\bE+\alpha
{\operatorname{Tr}}{\int}\widehat{\cal D}\bn\times \widehat{\bE}\,{\de^3 n}
\\
&\,i\hbar\left(\frac{\partial\tilde\rho}{\partial t}+{\operatorname{div}}(\bu\tilde\rho)+\frac{\bL}{JD}\cdot\mathscr{D}_\bn\tilde\rho\right)
=
\bigg[\widehat{H}_Q-\alpha\bn\cdot\widehat{\bE}
-\frac{\alpha}{\tilde{D}}
\Big[\mathscr{D}_{\bn}\tilde\rho-\frac12\tilde\rho\mathscr{D}_{\bn}\tilde{D},\cdot\widehat{\bE}\Big],\tilde\rho\bigg]
\label{MarcusHyd2PolA},
\end{align}
where we recognize the electric torques balancing the angular momentum equation, while the quantum evolution carries the backreaction terms on the right hand side of \eqref{MarcusHyd2PolA}. Together with the mass transport equation \eqref{EhrKoopHam4Pol}, the equations above provide a nonlocal dielectric model of ideal solvation hydrodynamics.

Given the intricate nonlocal form of the dielectric solvation model above, here we will look at the simplified case given by Marcus' \emph{local approximation} \cite{BuBa06,ChBa91,Marcus}. The latter consists in writing $\nabla\operatorname{div}\Delta^{-1}=\boldsymbol{1}+\Delta^{-1}{\operatorname{curl}}^2$ and neglecting the transverse polarization components $\operatorname{curl}\bP$, so that $\nabla\operatorname{div}\Delta^{-1}\simeq\boldsymbol{1}$ and ${\widehat{\cal H}=\widehat{H}_Q+\alpha\zeta\bn\cdot\widehat{\bmu}/\epsilon_0}$. Then, the solvation energy becomes
\begin{multline*}
{\sf h}={\int}\bigg(\frac1{2}\frac{m^2}{MD}+\frac1{2}\frac{L^2}{JD}+D\mathscr{U}(D) +\frac12(D-{D}_e)C{\ast}(D-{D}_e)
\\
+ \frac{|\bP|^2}{2\chi_0}
+
{\operatorname{ReTr}}{\int}\bigg(\tilde\rho\widehat{H}_Q+\frac{\alpha\zeta}{\epsilon_0}\tilde\rho \bn\cdot\widehat{\bmu}+{i\hbar}\frac{\alpha\zeta}{\epsilon_0\tilde{D}}\tilde\rho\{\tilde\rho,\bn\cdot\widehat{\bmu}\}_{S^2\!}\bigg){\de^3n}
\bigg) \de^3q.
\end{multline*}
In this case, the equations simplify considerably since the local approximation takes the Kelvin polarization force into a simple pressure term, while the solute backreaction term no longer carries the electrostatic kernel. In particular, 
 the system \eqref{EhrKoopHam1Pol}-\eqref{EhrKoopHam4Pol} specializes to
\begin{align}\label{MarcusHyd1Pol}
&MD\bigg(\frac{\partial}{\partial t}+\bu\cdot\nabla\bigg)\bu=-\nabla\bigg({\sf p}+\frac{|\bP|^2}{2\chi_0}\bigg)-{D}\nabla C{*}(D-D_e)
-\frac{\alpha}{\epsilon_0}
{\operatorname{Tr}}{\int}\widehat{\cal D}(\bn\cdot\widehat\bmu){\de^3 n}\,\nabla\zeta
\\\label{MarcusHyd2Pol}
&\,\frac{\partial \bL}{\partial t}+\operatorname{div}(\bu\bL)=-\frac{\alpha\zeta}{\epsilon_0}
{\operatorname{Tr}}{\int}\widehat{\cal D}\bn\times \widehat{\bmu}\,{\de^3 n}
\\\label{MarcusHyd3Pol}
&\,i\hbar\left(\frac{\partial\tilde\rho}{\partial t}+{\operatorname{div}}(\bu\tilde\rho)+\frac{\bL}{JD}\cdot\mathscr{D}_\bn\tilde\rho\right)
=
\bigg[\widehat{H}_Q+\frac{\alpha\zeta}{\epsilon_0}\bn\cdot\widehat{\bmu}
+\frac{\alpha\zeta}{\epsilon_0\tilde{D}}
\Big[\mathscr{D}_{\bn}\tilde\rho-\frac12\tilde\rho\mathscr{D}_{\bn}\tilde{D},\cdot\widehat{\bmu}\Big],\tilde\rho\bigg]
\\&\label{MarcusHyd4Pol}
\frac{\partial D}{\partial t}+\operatorname{div}\left(\bu D\right)=0.
\end{align}
The main feature of the local approximation is that in this case the solvent motion, in both its translational and orientational parts, is essentially produced by the  solute-solvent interaction, which becomes the driving feature of the solvation system. Based on Marcus' local approximation, the equations \eqref{MarcusHyd1Pol}-\eqref{MarcusHyd4Pol} provide a \emph{Marcus hydrodynamics} model of ideal solvation. Notice that, in this case, the equations  contain only first-order gradients of the dynamical variables. Similarly to the model in \S\ref{sec:NOEx}, here the barotropic-type pressure acts as the thermodynamic anchor of the solvent, while the  solvent-solvent Coulomb potential resolves the long-range polarization energy and structural interactions required for the dynamic inertial Marcus theory.
The microscopic packing and excluded-volume effects are included into the treatment by the autocorrelation function $C(\bq)$, as in \S\ref{sec:NOEx}.

A special role in the system \eqref{MarcusHyd1Pol}-\eqref{MarcusHyd4Pol} is played by the pseudo-density operator $\widehat{\cal D}$, whose parenthesis terms in  \eqref{pseudoD} render the extra backreaction beyond Ehrenfest dynamics. Indeed, as already shown in the context of phase-space dynamics \cite{BaBeGBTr24}, the corresponding terms are responsible for capturing backrection effects that cannot be retained by the standard Ehrenfest approach. For example, as explained in \cite{GBTr-fluid}, the parenthesis terms in \eqref{pseudoD} ensure that the classical solvent motion does not decouple from solute dynamics in the case of pure-dephasing dynamics such as $\widehat{\bmu}=\mu\widehat{\sigma}_3\mathbf{e}_3$. In more generality, this type of backreaction  captures decoherence with accuracy levels well beyond Ehrenfest dynamics  \cite{BaBeGBTr24}.

\section{Dissipative effects and polarization diffusion\label{sec:dissip}}

Having illustrated the ideal inertial regime of solvation hydrodynamics, we now move on to complete the discussion by restoring dissipative effects. Upon focusing on Marcus solvation, we start by adding standard momentum drags and viscous contributions in such a way that the equations \eqref{MarcusHyd1Pol}-\eqref{MarcusHyd2Pol} and \eqref{MarcusHyd4Pol}  become
\begin{align}\nonumber
&MD\bigg(\frac{\partial}{\partial t}+\bu\cdot\nabla\bigg)\bu=-\nabla\bigg({\sf p}+\frac{|\bP|^2}{2\chi_0}\bigg)-\nabla C{*}(D-D_e)
-\frac{\alpha}{\epsilon_0}
{\operatorname{Tr}}{\int}\widehat{\cal D}(\bn\cdot\widehat\bmu){\de^3 n}\,\nabla\zeta
\\\label{MarcusHyd1PolDiss}
&\hspace{5.5cm}+\eta\,{\operatorname{div}}\bigg(\nabla \bu+(\nabla\bu)^T-\frac23(\operatorname{div}\bu)\boldsymbol{1}\bigg)-\eta'\Delta\bu-\xi\bu
\\\label{MarcusHyd2PolDiss}
&\,\frac{\partial \bL}{\partial t}+\operatorname{div}(\bu\bL)=-\frac{\alpha\zeta}{\epsilon_0}
{\operatorname{Tr}}{\int}\widehat{\cal D}\bn\times \widehat{\bmu}\,{\de^3 n}-\gamma\bL
\\&\label{MarcusHyd4PolDiss}
\frac{\partial D}{\partial t}+\operatorname{div}\left(\bu D\right)=0.
\end{align}
Here, the coefficients $\eta$ and $\eta'$ are the shear and bulk viscosity, respectively, while $\xi$ is the solvent friction coefficient. Also, $\gamma$ is a rotational friction coefficient. Notice that here we refrain from considering mass diffusion in equation \eqref{MarcusHyd4PolDiss}.

In addition to these conventional dissipative terms, we incorporate the rotational dissipation of the polar solvent. Here, we specifically focus on the polarization relaxation underlying dielectric diffusion \cite{BaBi99,BaJa10}. Unlike angular momentum viscosity, polarization diffusion directly parametrizes the non-local spatial correlations of the solvent, ensuring that the solvation structure is resolved at physically realistic molecular lengthscales. While rotational viscosity offers a rigorous hydrodynamic description of torque dissipation, the empirical evaluation of its corresponding parameter remains challenging for molecular solvents. In contrast, the polarization diffusion coefficient can be  estimated from the experimental correlation length of the liquid. 

However, incorporating polarization diffusion represents a major modeling challenge because, {unlike other models \cite{BuBa06,ChBa91},} our present treatment does not carry a separate equation for the solvent polarization $\bP$, but rather the latter is computed as $\bP=\alpha{\operatorname{Tr}}{\int}\tilde\rho\bn{\de^3 n}$ using the quantum-classical density operator $\tilde\rho(\bq,\bn)$. This means that the only way to render dielectric diffusion effects consists in adding a suitable dissipative term to the quantum-classical evolution equation \eqref{MarcusHyd3Pol}. However, such a term must have the specific property of affecting only the orientational dynamics while, for example, avoiding the introduction of mass diffusion which is not considered in the present treatment. In addition, the extra dissipative term must only add diffusion to the classical orientational dynamics without affecting the motion of the quantum solute state $\varrho={\int}\tilde\rho\de^3n\de^3q$. In order to satisfy these criteria, here we propose adding the following term to the right-hand side of \eqref{MarcusHyd3Pol}:
\beq\label{DiffEntropy}
\nu\operatorname{div}\left({\tilde\rho}\operatorname{Tr}\bigg(\frac{\tilde\rho}{\tilde{D}}\nabla\frac{\delta S}{\delta \tilde\rho}\bigg)\right)
,\qquad\text{ where }\qquad
S=-{\int} \left({\int}\tilde{D}\ln\tilde{D}\,{\de^3 n} -D\ln D\right)\de^3q.
\eeq
Here, $S$ is the the orientational entropy given by the difference between the overall classical entropy and that associated to translational motion. This form of operator is inspired by the structure of dynamical density-functional theory models \cite{ChBa91b,GoNoKa16} as well as previous work by the authors on dissipative brackets for porous media flows \cite{HoPuTr10,HoPuTr08}. Dissipative brackets have a long history \cite{Gr84,Ka1984,Mo1984} and are widely used in several areas as a structure-preserving modeling tool in dealing with irreversible effects. 

Evaluating the {functional} derivative in \eqref{DiffEntropy} yields
\[
\nu\operatorname{div}\left({\tilde\rho}\operatorname{Tr}\bigg(\frac{\tilde\rho}{\tilde{D}}\nabla\frac{\delta S}{\delta \tilde\rho}\bigg)\right)=\nu\operatorname{div}\left(\frac{\tilde\rho}{f}\nabla f\right)
,\qquad\text{ where }\qquad
f=\frac{\tilde{D}}D
\]
is the conditional probability of the orientation  $\bn$ given the position $\bq$, so that ${\int}f(\bq,\bn){\de^3 n}=1$ at all points in space. 
In order to see the effect of this operator on averages, we let $A=A(\bq)$ be a classical solvent observable  and compute:
\[
\nu{\operatorname{div Tr}}{\int}\left(A\frac{\tilde\rho}{f}\nabla f\right){\de^3 n}=
\nu{\operatorname{div}}{\int}\bigg(\frac{\tilde{D} A}{f}\nabla f\bigg){\de^3 n}
=
\nu{\operatorname{div}}\left({D A}\nabla {\int}f{\de^3 n}\right)=0,
\]
so that no diffusion is added to solvent observables that are independent of the orientational coordinate. Instead, if we let $A=A(\bn)$ be an orientational solvent observable, we have
\[
\nu{\operatorname{div Tr}}{\int}\left(A\frac{\tilde\rho}{f}\nabla f\right){\de^3 n}=
\nu{\operatorname{div}}{\int}\bigg(\frac{\tilde{D} A}{f}\nabla f\bigg){\de^3 n}
=
\nu{\operatorname{div}}\left({D}\nabla{\int}f A {\de^3 n}\right).
\]
For example, letting $A(\bn)=\alpha\bn$ leads to the polarization relaxation term
\[
\nu{\operatorname{div}}\left(D\nabla\frac{ \bP}D\right),
\]
which reduces to the usual linear diffusion when $D$ is spatially constant.

With the addition of this diffusion term, the quantum-classical evolution equation \eqref{MarcusHyd3Pol} becomes
\beq\label{MarcusHyd5PolDiss}
i\hbar\left(\frac{\partial\tilde\rho}{\partial t}+{\operatorname{div}}(\tilde\rho\bu-\nu\tilde\rho\nabla {\ln} f)+\frac{\bL}{JD}\cdot\mathscr{D}_\bn\tilde\rho\right)
=
\bigg[\widehat{H}_Q+\frac{\alpha\zeta}{\epsilon_0}\bn\cdot\widehat{\bmu}
+\frac{\alpha\zeta}{\epsilon_0\tilde{D}}
\Big[\mathscr{D}_{\bn}\tilde\rho-\frac12\tilde\rho\mathscr{D}_{\bn}\tilde{D},\cdot\widehat{\bmu}\Big],\tilde\rho\bigg]
\eeq
Overall, the equations \eqref{MarcusHyd1PolDiss}-\eqref{MarcusHyd4PolDiss} and \eqref{MarcusHyd5PolDiss} comprise the dissipative model of Marcus solvation hydrodynamics. The insertion of the terms considered in this section can be also applied to the nonlocal dielectric solvation model to obtain a more realistic description of solvation hydrodynamics. However, given their intrinsic complexity, here we will refrain from presenting the corresponding equations.

\section{Conclusions}

Motivated by the necessity to solve the inertial solute-solvent interaction timescales in ultrafast solvation processes, this paper presented a quantum-classical modeling framework for nonadiabatic solvation hydrodynamics. Alleviating the computational complexity of previous similar approaches, this framework blends the Hamiltonian Poisson-bracket approach with recent advances in Koopman wavefunction methods and resorts to dynamical density-functional theory to render polarization diffusion. While the Hamiltonian approach ensures exact energy and momentum balance in the ideal conservative limit, the use of Koopman wavefunctions allows to capture backreaction and decoherence effects beyond Ehrenfest approaches while ensuring positivity of probability densities. {The latter is an important property whose enforcement requires special care: as we commented in \S\ref{sec:intro}, even small simple steps in the development of the closure model may destroy this and other basic physical properties}.  When the solvent polarization is included in the treatment, the combination of dynamical DFT with the dissipative brackets proposed in \cite{HoPuTr10,HoPuTr08} allows the selective diffusion of the solvent polarization in an entirely quantum-classical treatment while ensuring the unitary dynamics of the quantum solute state.

Largely inspired by the work of Burghardt and Bagchi \cite{BuBa06}, the treatment is based on the application of the moment method starting from a full phase-space approach. However, instead of using the established quantum-classical Liouville equation {\cite{Aleksandrov,Gerasimenko,Kapral}}, here we overcome its well-known drawbacks by resorting to the Koopman modeling strategy, in which the quantum backreaction carries its own associated energy in a completely Hamiltonian reversible setting. {As a result, the standard Ehrenfest model is extended} to account for electronic decoherence \cite{BaBeGBTr24,BeMaTr26}. In the Poisson-bracket approach presented here, one obtains a moment closure scheme by suitably expressing the backreaction energy functional in terms of relevant hydrodynamic variables. After the procedure is completed in the ideal conservative regime, dissipative effects are added \emph{a posteriori} to each relevant equation of motion.

The first part of the paper dealt with nonpolar solvation. After extending the conventional Ehrenfest hydrodynamics to account for adiabatic equations of state and the accompanying entropy transport, the treatment {moved} on to the Koopman model by providing a suitable closure for the backreaction energy functional. The resulting model was specialized to  the nonpolar solvation of nitric oxide in supercritical argon. Based on a quantum-classical Hamiltonian operator of pure-dephasing type, this problem is known to lead to quantum-classical decoupling for Ehrenfest dynamics \cite{GBTr-fluid,Manfredi23}. On the contrary, in the alternative treatment proposed here the model retains the coupling between the classical solvent and the quantum solute at all times.

In its second part, the paper extended the treatment to polar solvation hydrodynamics. While previous similar treatments involved moments on the extended configuration space retaining the orientational coordinate, in the present approach  all hydrodynamic variables are defined on physical space, with the only exception of the quantum-classical state variable. {In addition, the polar solvation model presented here does not lead to increasing the number of field variables beyond Ehrenfest dynamics.}
As a result, the methodology presented here alleviates computational complexity. Again, the Ehrenfest case was treated first and the Koopman approach followed thereafter. In the latter case, we neglected the backreaction energy associated to translational degrees of freedom while retaining the orientational polarization effects. In this way, spatial coordinates are treated by the usual Ehrenfest approach while the orientational backreaction is fully captured within the Koopman framework. We specialized the model to consider the problem of polar solvation in a fully nonlocal dielectric medium and then applied Marcus' local approximation, {thereby} neglecting the energy contributions of the transverse polarization components.

In the final section, the relevant dissipative terms were added to the ideal conservative equations following the standard procedure. An exception {to this standard methodology} was given by the addition of polarization diffusion, which required a careful blending of dynamical DFT approaches with dissipative brackets applying to porous media flows. As a result, we were able to formulate a quantum-classical diffusion term that directly affects the orientational solvent dynamics while avoiding mass diffusion and ensuring that the quantum solute dynamics retains its intrinsic unitary dynamics. As a final consideration, we note that the implementation of continuum models requires routine calibration of the internal fluid parameters against bulk solvent properties. Establishing these numerical connections via standard equilibrium molecular dynamics or ultrafast spectroscopic tracking remains a natural avenue for future studies.

\paragraph{Acknowledgments.} We are greatly indebted to Bartosz B\l{}asiak, Irene Burghardt, Francesco Di Maiolo, Raymond Kapral, Byung Kyu Na, and Klaus Schleich for several enlightening discussions on this and related topics. Also, CT is grateful to the Division of Physical and Mathematical Sciences at NTU Singapore for their kind hospitality during the course of this work.

\paragraph{Funding.} The authors acknowledge financial support by the Leverhulme Trust Research Project Grant RPG-2023-078. 


\appendix

\section{Quantum-classical Koopman equation in phase-space\label{sec:Appx}}
This appendix provides the details of the calculations taking to the quantum-classical equation of motions \eqref{KMod2} from its Hamiltonian form \eqref{KMod1}. {In addition, we show how the latter reduces to the Ehrenfest model equation \eqref{EhrPSeqn}}. After defining 
\beq\label{newvars}
\mathcal{D}={\operatorname{Tr}}\widehat{\cal P}
\,,\qquad\text{and}\qquad
\widehat{\sf P}=\frac{\widehat{\cal P}}{\mathcal{D}}
\eeq
and introducing the real-valued matrix pairing
$\langle\widehat{A},\widehat{B}\rangle={\operatorname{Re}}{\operatorname{Tr}}(\widehat{A}^\dagger\widehat{B})$,
it is convenient to consider the energy functional $h(\widehat{\mathcal{P}})$ \eqref{KMod1} in the form
\beq\label{AltHamForm}
h(\mathcal{D},\hat\varrho)=\int \!\mathcal{D}\Big\langle\hat\varrho,\widehat{H}+\frac{i\hbar}2[\nabla\widehat{\sf P},\bX_{\widehat{H}}]\Big\rangle\,\de q\de p=:\int \!\mathcal{D}\varepsilon(\widehat{\sf P},\nabla\widehat{\sf P},{\widehat{H}},\nabla {\widehat{H}})\,\de q\de p,
\eeq
where we recall the notation $\bX_{\widehat{A}}=\Bbb{J}\nabla\widehat{A}$, with $\Bbb{J}^{k\ell}=\{z^k,z^\ell\}$ and $\bz=(q,p)$.

\smallskip
\noindent{\bf The transport vector field in phase-space.}
As a first step, here we deal with the explicit form of the transport vector field $\langle \bX_{\delta h/\delta\widehat{\cal P}}\rangle$. The functional chain rule leads to rewriting
\begin{align*}
\big\langle X_{\delta h/\delta\widehat{\cal P}}^k\big\rangle
=&\,
\frac1{{\operatorname{Tr}}\widehat{\cal P}}\,\Bbb{J}^{k\ell}{\operatorname{Tr}}\bigg(\widehat{\cal P}\partial_\ell\frac{\delta h}{\delta \widehat{\cal P}}\bigg)
=
\frac1{\mathcal{D}}
\Bbb{J}^{k\ell}\bigg(\mathcal{D}\partial_\ell \frac{\delta h}{\delta \mathcal{D}}-{\operatorname{Tr}}\bigg(\frac{\delta h}{\delta \widehat{\sf P}}\partial_\ell\widehat{\sf P}\bigg)\bigg)
\end{align*}
and we further rearrange according to
\begin{align*}
&\,\mathcal{D}\partial_\ell \frac{\delta h}{\delta \mathcal{D}}-{\operatorname{Tr}}\bigg(\frac{\delta h}{\delta \widehat{\sf P}}\partial_\ell\widehat{\sf P}\bigg)
\\
=&
-\mathcal{D}\partial_\ell\varepsilon+\bigg\langle \mathcal{D}\frac{\partial \varepsilon}{\partial \widehat{\sf P}}-\operatorname{div}\Big(\mathcal{D}\frac{\partial \varepsilon}{\partial \nabla\widehat{\sf P}}\Big),\partial_\ell\widehat{\sf P}\bigg\rangle
\\\nonumber
=&
-\mathcal{D}\partial_\ell\varepsilon+\mathcal{D}\bigg\langle \frac{\partial \varepsilon}{\partial \widehat{\sf P}},\partial_\ell\widehat{\sf P}\bigg\rangle-\operatorname{div}\bigg\langle \mathcal{D}\frac{\partial \varepsilon}{\partial \nabla\widehat{\sf P}},\partial_\ell\widehat{\sf P}\bigg\rangle
+\mathcal{D}\bigg\langle \frac{\partial \varepsilon}{\partial \nabla\widehat{\sf P}},\nabla\partial_\ell\widehat{\sf P}\bigg\rangle
\\\nonumber
= &
-\mathcal{D}\partial_\ell\varepsilon+\mathcal{D}\bigg(\partial_\ell\varepsilon-\bigg\langle \frac{\partial \varepsilon}{\partial {\widehat{H}}},\partial_\ell {\widehat{H}}\bigg\rangle-\bigg\langle \frac{\partial \varepsilon}{\partial \nabla {\widehat{H}}},\partial_\ell\nabla {\widehat{H}}\bigg\rangle\bigg)-\operatorname{div}\bigg\langle \mathcal{D}\frac{\partial \varepsilon}{\partial \nabla\widehat{\sf P}},\partial_\ell\widehat{\sf P}\bigg\rangle
\\
= &
-\mathcal{D}\bigg(\bigg\langle \frac{\partial \varepsilon}{\partial {\widehat{H}}},\partial_\ell {\widehat{H}}\bigg\rangle+\bigg\langle \frac{\partial \varepsilon}{\partial \nabla {\widehat{H}}},\partial_\ell\nabla {\widehat{H}}\bigg\rangle\bigg)-\operatorname{div}\bigg\langle \mathcal{D}\frac{\partial \varepsilon}{\partial \nabla\widehat{\sf P}},\partial_\ell\widehat{\sf P}\bigg\rangle.
\end{align*}
At this point, upon recalling the notation $[\widehat{\bf A},\widehat{\bf B}]=[\widehat{A}_j,\widehat{B}_j]=\widehat{\bf A}\cdot\widehat{\bf B}-\widehat{\bf B}\cdot\widehat{\bf A}$, it is convenient to write
\beq
\varepsilon-\langle{\widehat{H}}\rangle
=
\frac\hbar2\big\langle i[\nabla\widehat{\sf P},\bX_{\widehat{H}}]\big\rangle
=
-\frac\hbar2\big\langle \nabla\widehat{\sf P},i[\widehat{\sf P},\bX_{\widehat{H}}]\big\rangle
=
\frac\hbar2\big\langle \bX_{\widehat{H}},i[\widehat{\sf P},\nabla\widehat{\sf P}]\big\rangle
=
-\frac\hbar2\big\langle \nabla {\widehat{H}},i[\widehat{\sf P},\bX_{\widehat{\sf P}}]\big\rangle,
\label{laura}
\eeq
so that
\[
\frac{\partial \varepsilon}{\partial \nabla\widehat{\sf P}}=-\frac{i\hbar}2[\widehat{\sf P},\bX_{\widehat{H}}]
\,,\qquad\quad
\frac{\partial \varepsilon}{\partial \nabla {\widehat{H}}}=-\frac{i\hbar}2[\widehat{\sf P},\bX_{\widehat{\sf P}}]
\,,\qquad\quad\text{and}\qquad\quad
\frac{\partial \varepsilon}{\partial  {\widehat{H}}}=\widehat{\sf P}.
\]
Thus, upon recalling the definition $\widehat\bGamma={i\hbar}\mathcal{D}[\widehat{\sf P},\nabla\widehat{\sf P}]/2$ in \eqref{somedefs}, we have
\begin{align*}
&\,-\mathcal{D}\bigg(\bigg\langle \frac{\partial \varepsilon}{\partial {\widehat{H}}},\partial_\ell {\widehat{H}}\bigg\rangle+\bigg\langle \frac{\partial \varepsilon}{\partial \nabla {\widehat{H}}},\nabla\partial_\ell {\widehat{H}}\bigg\rangle\bigg)-\operatorname{div}\bigg\langle \mathcal{D}\frac{\partial \varepsilon}{\partial \nabla\widehat{\sf P}},\partial_\ell\widehat{\sf P}\bigg\rangle
\\
&\,=
-\mathcal{D}\langle \partial_\ell {\widehat{H}}\rangle+\frac\hbar2\mathcal{D}\big\langle [i\widehat{\sf P},\bX_{\widehat{\sf P}}],\nabla\partial_\ell {\widehat{H}}\big\rangle+\frac\hbar2\operatorname{div}\big\langle \mathcal{D}[i\widehat{\sf P},\bX_{\widehat{H}}],\partial_\ell\widehat{\sf P}\big\rangle
\\\nonumber
&\,=
-\mathcal{D}\langle \partial_\ell {\widehat{H}}\rangle+\big\langle (\Bbb{J}\widehat\bGamma\cdot\nabla),\partial_\ell {\widehat{H}}\big\rangle+\frac\hbar2\operatorname{div}\big\langle \mathcal{D}\bX_{\widehat{H}},[-i\widehat{\sf P},\partial_\ell\widehat{\sf P}]\big\rangle
\\\nonumber
&\,=
-\mathcal{D}\langle \partial_\ell {\widehat{H}}\rangle+\big\langle (\Bbb{J}\widehat\bGamma\cdot\nabla),\partial_\ell {\widehat{H}}\big\rangle-\big\langle (\bX_{\widehat{H}}\cdot\nabla),\widehat\bGamma\big\rangle,
\end{align*}
where  the second step follows from the third equality in \eqref{laura}. Then, one obtains
\beq\label{transvectfield}
\big\langle \bX_{\delta h/\delta\widehat{\cal P}}\big\rangle=\langle\bX_{\widehat{H}}\rangle+\frac1{\mathcal{D}}
{\operatorname{Tr}}\big[\bX_{\widehat{H}},\widehat\bGamma\big]_\text{\!\bfi JL},
\eeq
where the notation was introduced in \eqref{somedefs}.

\smallskip
\noindent{\bf The quantum commutator.}
Proceeding further, here we show how expanding the commutator $[{\delta h}/{\delta\widehat{\cal P}},\widehat{\cal P}]$ finally leads to two explicit forms of equation \eqref{KMod1}. Using \eqref{newvars} and the chain rule, we have
{
\begin{equation*}
\left[\frac{\delta h}{\delta\widehat{\cal P}},\widehat{\cal P}\right]
= 
\left[\frac{\delta h}{\delta\widehat{\sf P}},\widehat{\sf P}\right]
= 
\left[\mathcal{D}\frac{\partial \varepsilon}{\partial \widehat{\sf P}}-\operatorname{div}\Big(\mathcal{D}\frac{\partial \varepsilon}{\partial \nabla\widehat{\sf P}}\Big),\widehat{\sf P}\right]
.
\end{equation*}
Then, since
\[
\frac{\partial \varepsilon}{\partial \widehat{\sf P}}=\frac{i\hbar}2[\nabla\widehat{\sf P},\bX_{\widehat{H}}]
\,,\qquad\text{and}\qquad
\frac{\partial \varepsilon}{\partial \nabla\widehat{\sf P}}=-\frac{i\hbar}2[\widehat{\sf P},\bX_{\widehat{H}}],
\]
equation \eqref{KMod1} reads explicitly
\[
i\hbar\frac{\partial \widehat{\cal P}}{\partial t}
+i\hbar{\operatorname{div}}\Big(\widehat{\cal P}\big(\langle\bX_{\widehat{H}}\rangle+{\mathcal{D}}^{-1}
{\operatorname{Tr}}\big[\bX_{\widehat{H}},\widehat\bGamma\big]_\text{\!\bfi JL}\big)\Big)
=\left[\widehat{{H}}+i\hbar\big[\bX_{\widehat{{H}}},\mathcal{D}^{-1}\nabla\hat{\mathsf{P}}+\hat{\mathsf{P}}\nabla \ln\sqrt{\mathcal{D}}\,\big]
,\widehat{\cal P}\right].
\]
We observe that this equation is an $\hbar$-correction of the Ehrenfest model, since both the transport vector field \eqref{transvectfield} and the skew-Hermitian generator in the above commutator involve correcting $\hbar$-terms to Ehrenfest dynamics. When these terms are omitted, the equation above reduces to the Ehrenfest model equation \eqref{EhrPSeqn}.

In order to show that equation \eqref{KMod1} reads in the form \eqref{KMod2}, we proceed by rearranging:
}
\begin{equation*}
\left[\mathcal{D}\frac{\partial \varepsilon}{\partial \widehat{\sf P}}-\operatorname{div}\Big(\mathcal{D}\frac{\partial \varepsilon}{\partial \nabla\widehat{\sf P}}\Big),\widehat{\sf P}\right]
= 
\mathcal{D}\left[\frac{\partial \varepsilon}{\partial \widehat{\sf P}},\widehat{\sf P}\right]+\mathcal{D}\left[\frac{\partial \varepsilon}{\partial \nabla\widehat{\sf P}},\nabla\widehat{\sf P}\right]-\operatorname{div}\left[\mathcal{D}\frac{\partial \varepsilon}{\partial \nabla\widehat{\sf P}},\widehat{\sf P}\right]
\end{equation*}
so that 
\begin{align}\nonumber
  &\,\mathcal{D}\left[\frac{\partial \varepsilon}{\partial \widehat{\sf P}},\widehat{\sf P}\right]+\mathcal{D}\left[\frac{\partial \varepsilon}{\partial \nabla\widehat{\sf P}},\nabla\widehat{\sf P}\right]-\operatorname{div}\left[\mathcal{D}\frac{\partial \varepsilon}{\partial \nabla\widehat{\sf P}},\widehat{\sf P}\right]
\\\nonumber
= &\,
\mathcal{D}\left[H+\frac{i\hbar}2[\nabla\widehat{\sf P},\bX_{\widehat{H}}],\widehat{\sf P}\right]-\frac{i\hbar}2\mathcal{D}\left[[\widehat{\sf P},\bX_{\widehat{H}}],\nabla\widehat{\sf P}\right]+\frac{i\hbar}2{\operatorname{div}}\big[\mathcal{D}[\widehat{\sf P},\bX_{\widehat{H}}],\widehat{\sf P}\big]
\\\nonumber
= &\,
\mathcal{D}\left[H,\widehat{\sf P}\right]+\big[\bX_{\widehat{H}},\widehat\bGamma\big]-\frac{i\hbar}2\operatorname{div}\!\big(\mathcal{D}\left[\widehat{\sf P},[\widehat{\sf P},\bX_{\widehat{H}}]\right]\!\big)
\\\nonumber
= &\,
[H,\widehat{\cal P}]+\big[\bX_{\widehat{H}},\widehat{\bGamma}\big]+{i\hbar}\operatorname{div}\!\big(\langle\bX_{\widehat{H}}\rangle\widehat{\cal P}-(\widehat{\cal P}\bX_{\widehat{H}})^{\sf H}\big),
\label{rearrangment}
\end{align}
where we have used $\widehat{\mathcal{P}}(q,p)=\Upsilon(q,p)\Upsilon(q,p)^\dagger$ and used the Jacobi identity in the second equality.
Putting things together, we conclude that the equation \eqref{KMod1} recovers \eqref{KMod2}.

\section{Retaining  translational and rotational backreaction\label{app:translbackr}}
This appendix shows how to restore the translational backreaction that was neglected in \S\ref{sec:hybPolKoop} and later sections, thereby going beyond the assumption in \eqref{notranslapprox}.  Following the steps in  \S\ref{sec:KoopHyd1}, here we replace
\beq\label{translapprox}
{\operatorname{ReTr}}{\int} {i\hbar}{\cal F}\widehat{\sf P}\partial_\bq{\widehat{\cal H}}\cdot\partial_\bp\widehat{\sf P} 
\de^3q \de^3p\de^3\mu{\de^3n}
\simeq{\operatorname{ReTr}}
{\int}  \frac{i\hbar c}{\tilde{D}^2}\,\tilde\rho(\nabla b\cdot\nabla\tilde\rho\times\nabla\widehat{\cal H})\,
\de^3q {\de^3n}
\eeq
where $\tilde{D}=\operatorname{Tr}\tilde\rho$ and, as before, 
 $b(\bq,t)$ is a transported function satisfying $\partial_t b+\bu\cdot\nabla b=0$. Also, $\tilde{c}=\tilde{c}(\bq,\bn)$ is an orientational density depending parametrically on the position coordinate. Since this quantity belongs to the same space as $\operatorname{Tr}\tilde\rho/D$, we extend the treatment from Section \ref{sec:KoopHyd1} by prescribing the equation
\beq\label{MP-ceqn}
\frac{\partial \tilde{c}}{\partial t}+\frac{\delta{\sf h}}{\delta \bm}\cdot\nabla \tilde{c}+
\frac{\partial}{\partial \bn}\cdot\left(\tilde{c}\frac{\delta \sf h}{\delta \bL}\times\bn\right)=0
\,.
\eeq
which indeed coincides with the equation for $\operatorname{Tr}\tilde\rho/D$ obtained by combining \eqref{HHam3Pol} and \eqref{HHam4Pol}. Here, the variable $\tilde{c}(q,\bn)$ ensures convergence of the associated backreaction energy term. A possible initial condition for $\tilde{c}$ which guarantees convergence of the backreaction energy is $\tilde{c}(\bq,\bn,0)=\operatorname{Tr}\tilde\rho(\bq,\bn,0)$. Then following the procedure in \S\ref{sec:KoopHyd1}, the antisymmetry requirement leads to writing the   Poisson bracket structure as 
\begin{multline}
\{f,k\}=\!{\int}\bm\cdot\left(\frac{\delta k}{\delta \bm}\cdot\nabla\frac{\delta f}{\delta \bm}-\frac{\delta f}{\delta \bm}\cdot\nabla\frac{\delta k}{\delta \bm}\right)\de ^3 q
-
{\int}D\left(\frac{\delta f}{\delta \bm}\cdot\nabla\frac{\delta k}{\delta D}-\frac{\delta k}{\delta \bm}\cdot\nabla\frac{\delta f}{\delta D}\right)\de ^3 q
\\
-{\operatorname{Tr}}{\int} \tilde\rho\left( \frac{i}\hbar\left[\frac{\delta k}{\delta \tilde\rho},\frac{\delta f}{\delta \tilde\rho}\right]
+\frac{\delta f}{\delta \bm}\cdot\nabla\frac{\delta k}{\delta \tilde\rho}
-\frac{\delta k}{\delta \bm}\cdot\nabla\frac{\delta f}{\delta  \tilde\rho}
-\bn\cdot\bigg(\frac{\delta f}{\delta \bL}\times\frac{\partial}{\partial \bn}\frac{\delta k}{\delta \tilde\rho}
-
\frac{\delta k}{\delta \bL}\times\frac{\partial}{\partial \bn}\frac{\delta f}{\delta \tilde\rho}\bigg)
\right)\de^3  q\de^2  n
\\
-
\int \! b\operatorname{div}\!\left(\frac{\delta f}{\delta \bm}\frac{\delta k}{\delta b}-\frac{\delta k}{\delta \bm}\frac{\delta f}{\delta b}\right)\de ^3 q
-
{\int} \bL\cdot\left( \frac{\delta k}{\delta \bL}\times\frac{\delta f}{\delta \bL}
+\frac{\delta f}{\delta \bm}\cdot\nabla\frac{\delta k}{\delta \bL}
-\frac{\delta k}{\delta \bm}\cdot\nabla\frac{\delta f}{\delta  \bL}
\right)\de^3  q
\\
-
{\int} \left(\tilde{c}\operatorname{div}\!\left(\frac{\delta f}{\delta \bm}\frac{\delta k}{\delta \tilde{c}}-\frac{\delta k}{\delta \bm}\frac{\delta f}{\delta \tilde{c}}\right)-\bn\cdot\bigg(\frac{\delta f}{\delta \bL}\times\frac{\partial}{\partial \bn}\frac{\delta k}{\delta \tilde{c}}
-
\frac{\delta k}{\delta \bL}\times\frac{\partial}{\partial \bn}\frac{\delta f}{\delta \tilde{c}}\bigg)\right)\de ^3 q{\de^3 n}.
\label{Ehr-brktPolbis}
\end{multline}
Once again, questions about the Jacobi identity may be raised. Luckily, this bracket is shown to fall within the category of Lie-Poisson brackets on semidirect-product Lie groups \cite{MaRa98,MaRaWe84}, thereby ensuring the Jacobi identity. The resulting Hamiltonian equations read
\begin{align}\nonumber
&\frac{\partial \bm}{\partial t}+\operatorname{div}\bigg(\frac{\delta \sf h}{\delta \bm}\bm\bigg)+\nabla\frac{\delta \sf h}{\delta \bm}\cdot\bm=\frac{\delta \sf h}{\delta b}\nabla b-D\nabla\frac{\delta \sf h}{\delta D}
\\
&\,\hspace{7.25cm}-\nabla\frac{\delta \sf h}{\delta \bL}\cdot\bL
-{\operatorname{Tr}}{\int}
\tilde\rho\nabla\frac{\delta \sf h}{\delta \tilde\rho}\,{\de^3 n}
-{\int}
\tilde{c}\nabla\frac{\delta \sf h}{\delta \tilde{c}}\,{\de^3 n},
\label{HHam1PolApp}
\\&\label{HHam2PolApp}
\frac{\partial \bL}{\partial t}+\operatorname{div}\bigg(\frac{\delta \sf h}{\delta \bm}\bL\bigg)=-\bL\times\frac{\delta \sf h}{\delta \bL}-{\operatorname{Tr}}{\int}\tilde\rho\bn\times\frac{\partial}{\partial \bn}\frac{\delta \sf h}{\delta \tilde\rho}\,{\de^3 n}
-{\int}\tilde{c}\bn\times\frac{\partial}{\partial \bn}\frac{\delta \sf h}{\delta \tilde{c}}\,{\de^3 n},
\\&\label{HHam3PolApp}
i\hbar\frac{\partial \tilde\rho}{\partial t}+i\hbar\operatorname{div}\bigg(\tilde\rho\frac{\delta \sf h}{\delta \bm}\bigg)+{i\hbar}\frac{\partial}{\partial \bn}\cdot\left(\tilde\rho\frac{\delta \sf h}{\delta \bL}\times\bn\right)=\left[\frac{\delta \sf h}{\delta \tilde\rho},\tilde\rho\right],
\\&\label{HHam4PolApp}
\frac{\partial D}{\partial t}+\operatorname{div}\left(\bu D\right)=0
\,,\qquad
\frac{\partial b}{\partial t}+\bu\cdot\nabla b=0
\,,\qquad
\frac{\partial \tilde{c}}{\partial t}+\frac{\delta{\sf h}}{\delta \bm}\cdot\nabla \tilde{c}+
\frac{\partial}{\partial \bn}\cdot\left(\tilde{c}\frac{\delta \sf h}{\delta \bL}\times\bn\right)=0.
\end{align}

With the closure in \eqref{translapprox} and recalling \eqref{KoopClosure} as well as the notation \eqref{Nambubkt}, 
we eventually write
\beq\label{KoopClosureApp}
{\int}\big\langle i\hbar\{\widehat{\cal P},\widehat{H}\}\big\rangle \de^3q \de^3p\de^3\mu{\de^3n}
={\operatorname{ReTr}}
{\int}  \frac{i\hbar}{\tilde{D}}\tilde\rho\left(\frac{ \tilde{c}}{\tilde{D}}
\{\tilde\rho,\widehat{\cal H}\}_b+\{\tilde\rho,\widehat{\cal H}\}_{S^2}
\right)
\de^3q {\de^3n},
\eeq
Then, upon denoting $\hat\varrho=\tilde\rho/\tilde{D}$ and
\[
{\cal B}_T={\operatorname{ReTr}}
{\int}  \frac{i\hbar\tilde{c}}{\tilde{D}^2}\tilde\rho
\{\tilde\rho,\widehat{\cal H}\}_b\,
\de^3q {\de^3n}=:{\int}\varepsilon_T(\tilde{c},\hat\varrho,\nabla\varrho,\nabla\widehat{\cal H})\de^3q {\de^3n}
,
\]
we compute
\begin{align*}
&-{\operatorname{Tr}}
{\int}\tilde\rho\mathscr{D}_\bn\frac{\delta {\cal B}_T}{\delta \tilde\varrho}{\de^3n}
-{\int}\tilde{c}\mathscr{D}_\bn\frac{\delta {\cal B}_T}{\delta \tilde{c}}{\de^3n}
\\
=&\,{\operatorname{Tr}}
{\int}\frac{\delta {\cal B}_T}{\delta \hat\varrho}\mathscr{D}_\bn\hat\varrho{\de^3n}
-{\int}\tilde{c}\mathscr{D}_\bn\frac{\delta {\cal B}_T}{\delta \tilde{c}}{\de^3n}
\\
=&\,
{\operatorname{Tr}}
{\int}\left(\frac{\partial \varepsilon_T}{\partial \hat\varrho}-{\operatorname{div}}\frac{\partial \varepsilon_T}{\partial \nabla\hat\varrho}\right)\mathscr{D}_\bn\hat\varrho{\de^3n}
+{\int}\frac{\partial \varepsilon_T}{\partial \tilde{c}}\mathscr{D}_\bn \tilde{c}{\de^3n}
\\
=&\,
-\operatorname{div}
{\operatorname{Tr}}
{\int}\left(\frac{\partial \varepsilon_T}{\partial \nabla\hat\varrho}\mathscr{D}_\bn\hat\varrho
+
\frac{\partial \varepsilon_T}{\partial \nabla\widehat{\cal H}}\mathscr{D}_\bn\widehat{H}
\right){\de^3n}
+
{\operatorname{Tr}}
{\int}\bigg(\partial_\bn\cdot\frac{\partial \varepsilon_T}{\partial \partial_\bn\widehat{\cal H}}\bigg)\mathscr{D}_\bn\widehat{\cal H}{\de^3n}
\\
=&\,
-\frac12
{\operatorname{Tr}}
{\int}\big(\{i\hbar \tilde{c}[\hat\varrho,\mathscr{D}_\bn\hat\varrho],\widehat{\cal H}\}_b
-
\{i\hbar \tilde{c}[\hat\varrho,\mathscr{D}_\bn\widehat{\cal H}],\hat\varrho\}_b
\big){\de^3n}
-
{\operatorname{ReTr}}
{\int}i\hbar\{\tilde{c}\hat\varrho,\hat\varrho\}_{b}\mathscr{D}_\bn\widehat{\cal H}{\de^3n}.
\end{align*}
Thus, upon defining the quantum-classical pseudo-density operator 
\[
\widehat{\cal D}=\tilde\rho+\frac{i\hbar}2\bigg(\bigg\{\tilde\rho,\frac{\tilde\rho}{D\tilde{c}}\bigg\}_{b}+\bigg\{\tilde\rho,\frac{\tilde\rho}{\tilde{D}}\bigg\}_{\!S^2}-\bigg\{\frac{\tilde\rho}{D\tilde{c}},\tilde\rho\bigg\}_{b}-\bigg\{\frac{\tilde\rho}{\tilde{D}},\tilde\rho\bigg\}_{\!S^2}\bigg)
\]
for compactness of notation, using the total energy
\[
{\sf h}={\int}\left(\frac1{2M}\frac{m^2}{D}+\frac1{2J}\frac{L^2}{D}+D\mathscr{E}(D,\bP) + {\operatorname{Tr}}{\int}\widehat{\cal D}\widehat{\cal H}{\de^3 n}\right) \de^3q
\]  
in the equations \eqref{HHam1PolApp}-\eqref{HHam4PolApp} leads to
\begin{align}\label{EhrKoopHam1PolApp}
&MD\left(\frac{\partial}{\partial t}+\bu\cdot\nabla\right)\bu=-\nabla{\sf p}-\nabla\left(D\frac{\partial \mathscr{E}}{\partial \bP}\right)\cdot\bP
-{\operatorname{Tr}}{\int}\widehat{\cal D}\nabla  \widehat{\cal H}\,
{\de^3 n}
\\
&\,
\frac{\partial \bL}{\partial t}+\operatorname{div}(\bu\bL)=D\frac{\partial \mathscr{E}}{\partial \bP}\times\bP
\nonumber
\\&\,
\hspace{2.15cm}
-\frac12
{\operatorname{Tr}}
{\int}i\hbar\big(\{ \tilde{c}[\hat\varrho,\mathscr{D}_\bn\hat\varrho],\widehat{\cal H}\}_b
-
\{ \tilde{c}[\hat\varrho,\mathscr{D}_\bn\widehat{\cal H}],\hat\varrho\}_b
\big){\de^3n}
-{\operatorname{Tr}}{\int}\widehat{\cal D}\mathscr{D}_\bn \widehat{\cal H}\,{\de^3 n}
\label{EhrKoopHam42PolApp}
\\
&\,i\hbar\left(\frac{\partial}{\partial t}+\bu\cdot\nabla+\frac{\bL}{JD}\cdot\bn\times\frac{\partial}{\partial\bn}\right)\hat\varrho
=[\widehat{\cal H},\hat\varrho]
\nonumber
\\&\,\hspace{3.5cm}
+\left[\frac{c}{\tilde{D}}\left[\nabla\hat\varrho+\hat\varrho\nabla {\ln\sqrt{c}},\bX_{\widehat{\cal H}}^{(b)}\right]+\left[ \frac{\partial\hat\varrho}{\partial\bn}+\hat\varrho\frac{\partial}{\partial\bn} {\ln\sqrt{\tilde{D}}},\bX_{\widehat{\cal H}}^{(S^2)}\right]\!,\hat\varrho\right]
\label{EhrKoopHam3PolApp}
\\&\label{EhrKoopHam4PolApp}
\frac{\partial D}{\partial t}+\operatorname{div}\left(\bu D\right)=0,
\qquad\ 
\frac{\partial b}{\partial t}+u\cdot\nabla b=0,
\qquad\ 
\frac{\partial\tilde{c}}{\partial t}+\bu\cdot\nabla\tilde{c}+\frac{\bL}{JD}\cdot\bn\times\frac{\partial\tilde{c}}{\partial\bn}=0.
\end{align}
Here, we have denoted by
\[
\bX_{\widehat{\cal H}}^{(b)}=\nabla{\widehat{\cal H}}\times\nabla b
\,,\qquad\qquad
\bX_{\widehat{\cal H}}^{(S^2)}=\bn\times\frac{\partial\widehat{\cal H}}{\partial\bn},
\]
the Nambu and Lie-Poisson vector field on the sphere, respectively. Despite its underlying Hamiltonian structure, the system \eqref{EhrKoopHam1PolApp}-\eqref{EhrKoopHam4PolApp} has evidently a formidable level of complexity.

\end{document}